\documentclass[amsmath,amssymb,reprint,twocolumn, prb]{revtex4-1}

\usepackage[utf8]{inputenc}
\usepackage[T1]{fontenc}

\usepackage[english]{babel}

\usepackage{amsmath,amssymb,amsfonts}
\usepackage{amstext, mathrsfs, textcomp}
\usepackage{mathptmx}
\usepackage{bigints}  
\usepackage{nicefrac}
\usepackage{multirow}
\usepackage{dcolumn}
\usepackage{bm}

\usepackage{subfigure}
\usepackage{graphicx}
\usepackage{xcolor}
\usepackage[export]{adjustbox}

\usepackage[colorlinks=true, citecolor=blue, urlcolor=blue, linkcolor=blue]{hyperref}

\usepackage{changes}

\graphicspath{{Figures/}{TOC_ACS/}}

\newcommand{\FIGCAPTIONPREFIX}{}

\newcommand{\TITLE}{Deep learning meets nanophotonics: A generalized accurate predictor for near fields and far fields of arbitrary 3D nanostructures}

\begin{document}

	\title{\TITLE}
	
	\author{\firstname{Peter R.} \surname{Wiecha}}
	\email[e-mail~: ]{p.wiecha@soton.ac.uk}
	\affiliation{Physics and Astronomy, Faculty of Engineering and Physical Sciences, University of Southampton, SO 17 1BJ Southampton, UK}
	
	\author{\firstname{Otto L.} \surname{Muskens}}
	\email[e-mail~: ]{o.muskens@soton.ac.uk}
	\affiliation{Physics and Astronomy, Faculty of Engineering and Physical Sciences, University of Southampton, SO 17 1BJ Southampton, UK}

	\begin{abstract}
		Deep artificial neural networks are powerful tools with many possible applications in nanophotonics. Here, we demonstrate how a deep neural network can be used as a fast, general purpose predictor of the full near-field and far-field response of plasmonic and dielectric nanostructures. A trained neural network is shown to infer the internal fields of arbitrary three-dimensional nanostructures many orders of magnitude faster compared to conventional numerical simulations. Secondary physical quantities are derived from the deep learning predictions and faithfully reproduce a wide variety of physical effects without requiring specific training. We discuss the strengths and limitations of the neural network approach using a number of model studies of single particles and their near-field interactions. Our approach paves the way for fast, yet universal methods for design and analysis of nanophotonic systems.
		{\newline\textbf{keywords:} deep learning, nano-photonics, rapid nano-optics simulations, silicon nanostructures, plasmonics}
	\end{abstract}

	\maketitle

	Dielectric and metallic nanostructures of sub-wavelength size can be designed such that their interaction with light differs significantly from bulk materials. Nanophotonics aims to exploit optical resonances and strong localized fields that can be designed by nanoparticle geometry and material choices.\cite{draine_discrete-dipole_1994, evlyukhin_multipole_2011, kinkhabwala_large_2009, albellaLowLossElectricMagnetic2013} The unique nanoscale properties can be used in applications such as plasmonic nanoantennas for near-field energy concentration and meta-surfaces for directing and controlling light. More complex optical behavior can be designed such as polarization conversion,\cite{wiecha_polarization_2017} chirality, \cite{schaferling_tailoring_2012} localized heat generation \cite{baffou_thermoplasmonics_2010} or nonlinear optical effects \cite{kauranen_nonlinear_2012, shcherbakov_enhanced_2014, wiecha_origin_2016}.
	
	The interaction of light with many types of nanostructures can be modelled accurately by solving the classical Maxwell's equations,\cite{maxwell_dynamical_1865} for which many commercial and open source methods are available. However, numerical nanophotonic simulations are often time-consuming and can take hours or even days for complex systems.\cite{smajic_comparison_2009} Many subsequent simulations are needed for iterative optimization methods such as the rational design of nanosystems by topology optimization and inverse design.\cite{liu_training_2018,feichtner_evolutionary_2012, wiecha_evolutionary_2017, kingston_assessing_2019} Fast evaluation of optical properties is highly desirable also for example in real-time sensors based on 3D molecular nanorulers, \cite{liu_three-dimensional_2011, kuzyk_reconfigurable_2014} or in the active control of smart and reconfigurable nano-materials \cite{zhou_dna-nanotechnology-enabled_2017, qian_reversibly_2017}.
	
	Methods of artificial intelligence, and in particular deep learning,\cite{goodfellow_deep_2016, lecun_deep_2015} are powerful tools with potentially groundbreaking relevance for nano-optics. First captivating applications have been reported in experimental photonics and nano-optics. Examples are the possibility of phase recovery in conventional microscopy,\cite{rivenson_phase_2017} stabilization of lasers\cite{baumeister_deep_2018-1} and a large variety of applications in data analysis and interpretation\cite{cirovic_feed-forward_1997, ciresan_deep_2012, wang_artificial_2014, ziatdinov_deep_2017, jo_holographic_2017, zhang_analyzing_2018, han_deep_2019, wiecha_pushing_2019, timoshenko_probing_2019}. 
	For forward modelling, early work has shown that artificial neural networks (ANNs) can be used as approximate predictors for light-matter interaction phenomena or optical scattering at nanostructures. Examples are strong-field ionization of potassium atoms,\cite{selle_modelling_2008} SHG of a specific fluorescent molecule,\cite{selle_modeling_2007} the optical transmittance of ``H''-shaped particles\cite{malkiel_plasmonic_2018} or the scattering cross sections of multilayer spheres\cite{peurifoy_nanophotonic_2018}. All of those reported ANN techniques apply to single, very specific problems and for a particular nanostructure geometry.
	
	Furthermore, deep artificial neural networks are highly promising to approach notoriously difficult inverse problems in nano-optics, like the design of meta-surfaces or the tailoring of optical properties of individual nano-structures. Even though no generalized and fast inverse design method has as yet been reported, rapid progress has been made in this direction over the last two years.\cite{malkiel_plasmonic_2018, peurifoy_nanophotonic_2018, liu_generative_2018, liu_training_2018, liu_hybrid_2019, an_novel_2019, chen_smart_2019, liu_compounding_2019, jiang_global_2019} The overall trend in these studies so far is, that for every specific inverse problem using a particular geometric model, a neural network needs to be designed in a time demanding and computationally very expensive process, involving hyper-parameter optimization, training data generation, training and extensive testing. 
	
	\begin{figure*}[tb]
		\centering
		\includegraphics[page=1]{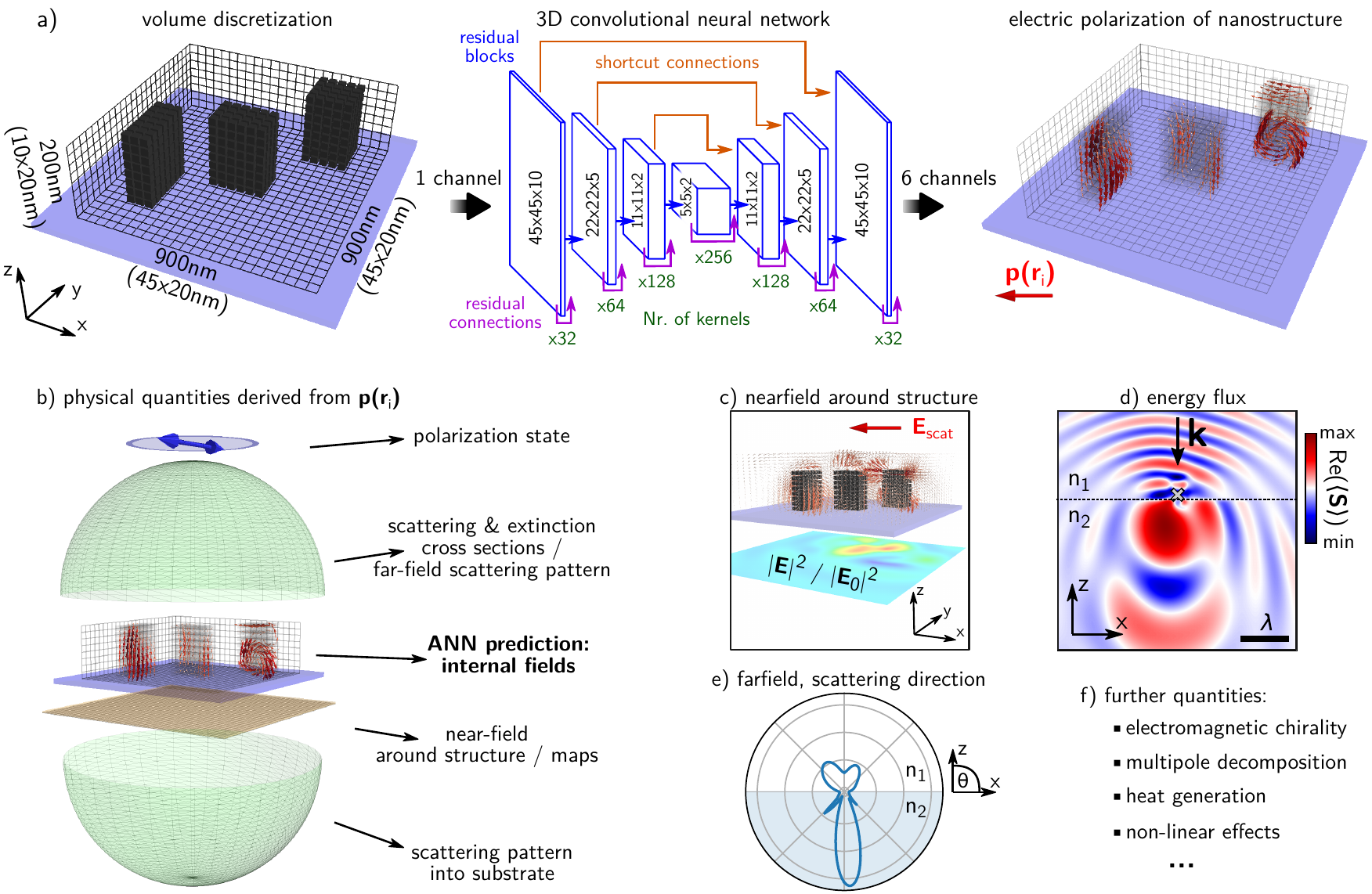}
		\caption{\FIGCAPTIONPREFIX
			(a) Sketch of the proposed neural network model for the example of the silicon nanostructure model.
			The volume discretization of the three-dimensional geometry (left, area of \(45 \times 45 \times 10\) meshpoints with \(20\,\)nm step, fixed height of \(200\,\)nm) is fed into the neural network.
			The three-dimensional convolutional network follows an encoder-decoder architecture and is organized in a sequence of residual blocks.
			The principal layout of these blocks, the number of kernels as well as the layer dimensions of the silicon predictor are shown in the sketch of the network.
			The six output channels of the network contain the real and imaginary parts of the \(x\), \(y\) and \(z\) components of the complex electric field inside the nanostructure.
			For more details see Section III of the Supporting Information.
			Following the calculation of the self-consistent electric polarization inside the structure, the latter can be interpreted as dipole moments \(\mathbf{p}(\mathbf{r}_i)\) of the single mesh cells at \(\mathbf{r}_i\). 
			(b) Various physical quantities in the near-field and far-field, as illustrated can be derived from \(\mathbf{p}(\mathbf{r}_i)\).
			This includes the electric or magnetic near-field (c), the Poynting vector (d) or far-field scattering patterns (e) among many others (f). The model includes a dielectric substrate (\(n_{\text{subst}}=1.45\)). The structure is illuminated by a linearly polarized plane wave from the top with \(\lambda_0 = 700\,\)nm.
		}
		\label{fig:overview}
	\end{figure*}

	Here, we present a general approach to nano-optical modeling which is distinct from all previous works by its capability for fast and accurate modeling of generalized nano-optical effects in a variety of nanostructures. Requiring only a single training procedure, our concept fully generalizes ANNs for nano-photonic simulations and allows to tremendously accelerate predictions for countless problems in nano-optics. We demonstrate that the generalized network captures a range of complex nano-optical near- and far-field effects in nanostructures, such as higher order antenna resonances, electric and magnetic dipole modes, non-radiating anapole states or Kerker-type directional scattering, without the need of any specific training for these effects. Our approach is based on a three-dimensional, fully convolutional neural network (CNN), trained on predicting a coupled dipole representation of the fields inside nanostructures of arbitrary shape. These predictions can be used subsequently to reconstruct many secondary physical quantities with uncertainties as low as few percent.

	\begin{figure*}[tb]
		\centering
		\includegraphics[page=1]{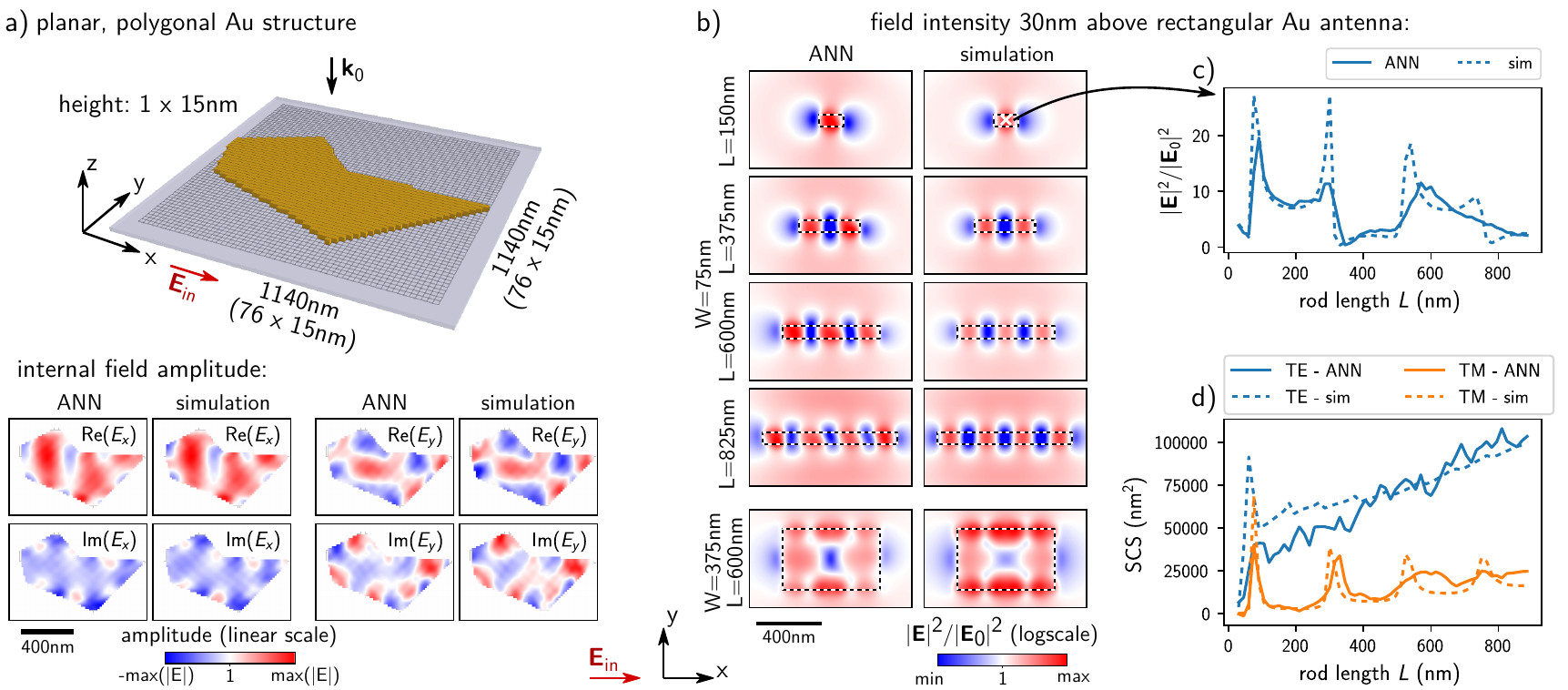}
		\caption{\FIGCAPTIONPREFIX
			Internal field prediction of polygonal, planar gold nanostructures.
			(a) Top: Sketch of the simulated planar gold structure. Discretization with a step of \(15\,\)nm on an area of \(76 \times 76\) positions with a height of a single layer of mesh-cells.
			The polygonal structure lies on a dielectric substrate (\(n_{\text{subst}}=1.45\)) and is illuminated by a linearly polarized plane wave from the top with \(\lambda_0 = 700\,\)nm.
			On the bottom, the real (top row) and imaginary parts (bottom row) of the \(E_x\) (two left columns) and \(E_y\) amplitude (two right columns) of the internal electric field is shown for ANN prediction and simulation. The linear and symmetric color scales are pair-wise normalized between ANN and simulation for each field component.
			More examples can be found in the SI, figure~S4.
			(b) Comparison of the electric field intensity \(30\,\)nm above rectangular gold antennas, calculated using the ANN predicted internal fields (left subplots) and by numerical simulations (right subplots). The colormap is on a symmetric, logarithmic scale, white corresponds to \(|\mathbf{E}|^2 = |\mathbf{E}_0|^2\). 
			(c) Electric field intensity above the center of a nanorod of \(W=75\,\)nm as function of the rod length (position indicated by white marker in the top right nearfield colorplot). Solid line: calculated using the ANN, dashed line: numerical simulation. 
			(d) scattering cross section (SCS) from ANN (solid lines) and GDM simulation (dashed lines) as function of rod length for an incident plane wave polarized either along \(Y\) (TE, blue) or along \(X\) (TM, orange). Scale bars in (a-b) are \(400\,\)nm.
		}
		\label{fig:planar_plasmonics}
	\end{figure*}

	An overview of the model description is shown in Figure~\ref{fig:overview}. The aim of our work is to develop an ANN capable of predicting the time-harmonic electric polarization density inside nanostructures of arbitrary shape. For our demonstration we use a geometric model consisting of a rectangular grid of positions, as illustrated by the example of Figure~\ref{fig:overview}a, left panel. Nanostructures are mapped onto the 3D grid using a standard volume discretization approach. The resulting geometry can be directly evaluated with numerical simulations using the coupled dipole approximation (CDA). Upon normal incidence plane wave illumination (\(\mathbf{k}\) along \(-Z\)) and linear polarization along \(X\), we numerically calculate the field at every mesh cell inside the nanostructure via the Green Dyadic Method (GDM),\cite{girard_near_2005} for which we use a home-built python implementation ``pyGDM''.\cite{wiecha_pygdmpython_2018} A short summary of the GDM formalism can be found in Section I of the Supporting Information. The model includes a dielectric substrate (\(n_{\text{subst}}=1.45\)).
	While in principle any simulation method can be used to generate the network training data, we use the GDM for its simplicity and its good convergence in the case of small nanostructures as well as for the fact that the results are by design available on the required cubic discretization grid.
	
	The conceptual beauty of the CDA method lies in the initial assumption that every meshpoint can be approximated by a dipolar polarizability, which allows to treat every cell inside the nanostructure as an oscillating dipole moment. In consequence, various physical quantities in the near-field as well as in the far-field region can be derived from the internal fields of a CDA simulation.
	Figure~\ref{fig:overview}b illustrates a few of the possible observables that can be calculated from the dipole discretization.
	Using according Green's dyads, the optical electric and magnetic fields (and in consequence the Poynting vector) can be obtained at any location outside the nanostructure (see Figure~\ref{fig:overview}(c-d)).\cite{wiecha_pygdmpython_2018}
	Extinction, scattering or absorption cross sections can be calculated almost effortless,\cite{draine_discrete-dipole_1994} as well as the polarization state and spatial patterns of the scattering (Figure~\ref{fig:overview}e),\cite{wiecha_polarization_2017, wiecha_strongly_2017} the dissipated heat or local temperature gradients,\cite{baffou_thermoplasmonics_2010, girard_designing_2018} nonlinear effects like second or third harmonic generation and multi-photon luminescence\cite{wiecha_origin_2016, balla_second_2010, teulle_scanning_2012} or the multipole decomposition of the optical response.\cite{evlyukhin_multipole_2011}
	In consequence, an ANN capable to accurately predict the internal electric fields of a photonic nanostructure represents a generalized, phenomenological model of light-matter interaction, with tremendous potentials for rapid nano-optical simulations.
	
	A 3D grid of positions is used as input layer to the network which is illustrated in Figure~\ref{fig:overview}a, middle panel. The network used in this study is a three-dimensional symmetric, fully convolutional network with ``U-Net''-type shortcut connections between corresponding convolutional and up-sampling units. This type of ANN is known to be particularly strong at the reconstruction of spatial information.\cite{ronneberger_u-net_2015} Shortcuts were found to be essential for our purpose to obtain good approximations of the internal fields. In addition to the U-Net design, we organize the network in residual blocks \cite{he_deep_2015, szegedy_inception-v4_2016} which allows us to maintain good learning performance on a very deep architecture with as much as 91~layers including 33~three-dimensional convolutional operations. All network details and hyperparameters are given in Section III of the Supporting Information.
	
	The output of the network is composed of \(6\) layers of the same size as the input grid, which correspond to the real and imaginary parts of \(E_x\), \(E_y\) and \(E_z\). We use complex amplitudes for the fields, so phase information and retardation effects are included in the neural network predictions. An illustration of real part field-vectors is shown in Figure~\ref{fig:overview}a, right panel.
	
	For the network training, we simulate the internal fields of 30000 random nanostructures.
	We test the network on two datasets, the first contains planar gold nanostructures (\(15\,\)nm height) of random polygonal shapes.
	The second set is composed of silicon pillar structures (\(200\,\)nm height), consisting of one or more arranged cuboidal blocks. 
	Details about the geometric models are given in Section II of the Supporting Information.
	For convenience we use fixed height structures throughout this demonstration, but we want to emphasize that this is not an inherent limitation of the approach. On the contrary, variable height structures can be modeled without any further modifications and without significant loss of accuracy, as shown in the supporting information figure~S9.
	We fix the illumination conditions to normal incidence plane wave excitation at a wavelength of \(\lambda_0=700\,\)nm with linear polarization along \(X\).
	In consequence, the neural network is limited to predictions under the conditions chosen for the training data generation. Hence, in our approach a separate ANN needs to be trained if the structure model or material is modified and also for every illumination configuration, e.g. for every angle of an oblique incidence illumination.
	We note that under normal incidence, any linear polarization angle can be achieved by a rotation of the nanostructure. Fully arbitrary polarization states of the illumination can furthermore be obtained via superposition of perpendicular linear polarizations, as demonstrated in the supporting information Fig.~S7 for left circular polarization.
	
	In both cases of plasmonic and dielectric structures we use 28000 samples for training and the remaining 2000 structures for validation and benchmarking. 
	On an NVIDIA P6000 GPU, the training with the gold (silicon) dataset takes around 2 (10) minutes per epoch. We stop training after 100 epochs, longer training leads to no further improvement in validation accuracy (see Figure S10 in the Supporting Information). On the same GPU, the trained network delivers its prediction in around \(3\,\)ms for a planar gold structure and \(6\,\)ms for a silicon structure. On a third generation Intel i7 quad-core CPU (i7-3770) the predictions for the gold, respectively silicon structures take around \(53\)\,ms and \(250\)\,ms. 
	Depending on the chosen hardware platform, this is \(3\) to \(5\) orders of magnitude faster than the conventional simulations, which take seconds to minutes on the Intel i7 CPU. 
	The simulation time for the full datasets, running in parallel on two workstations, was approximately \(10\) days. We note that since the simulations are entirely independent, the data generation is a so-called ``embarrassingly parallel'' task that can be parallelized to any extent.
	The GDM simulations can potentially be accelerated by GPUs, but according to literature, less than one order of magnitude speed-up is expected for GPU-based LU decomposition.\cite{ozcanInvestigationPerformanceLU2012}
	

	\begin{figure*}[tb]
		\centering
		\includegraphics[page=1]{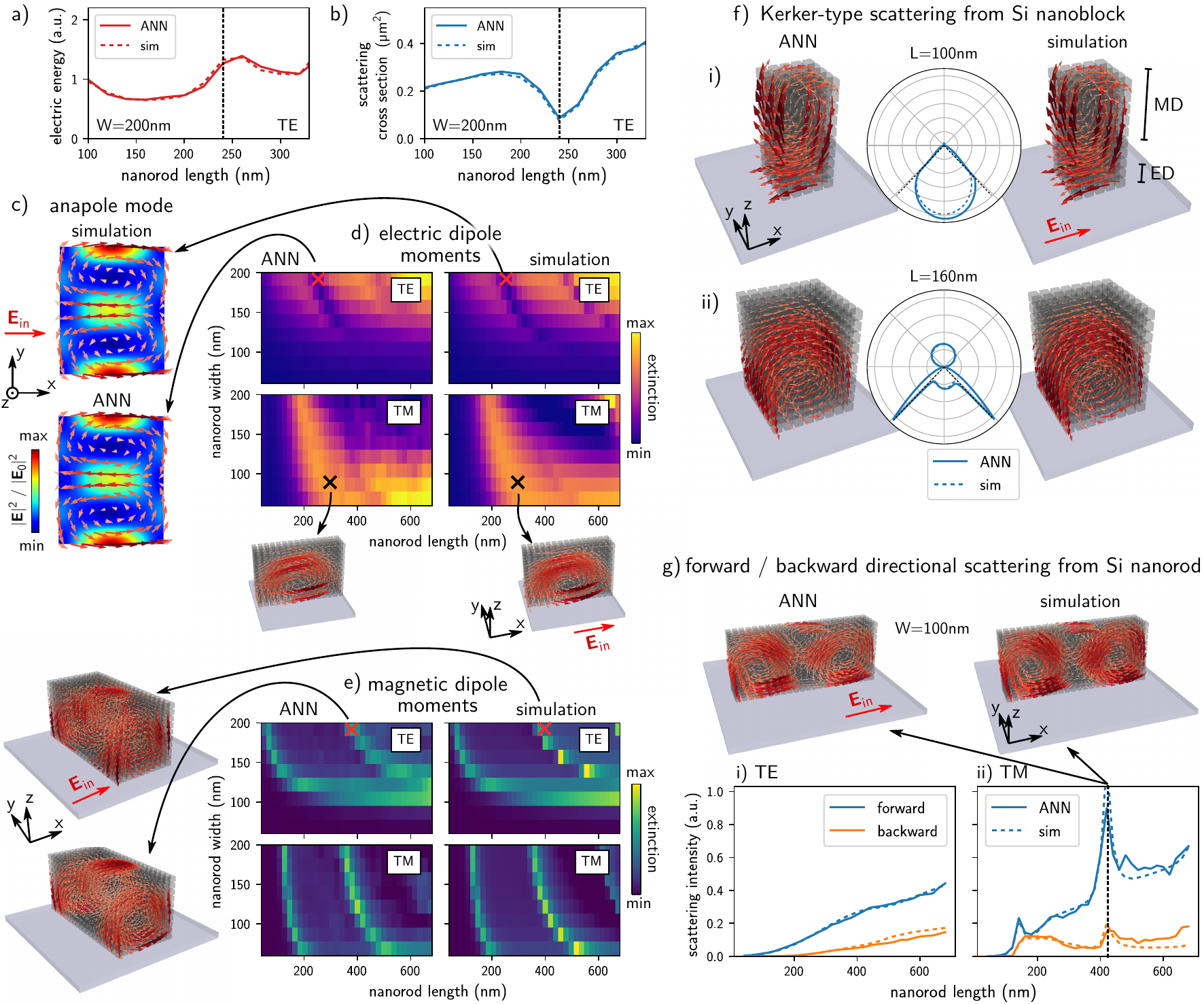}
		\caption{\FIGCAPTIONPREFIX
			Reproducibility of complex nano-optical effects by the ANN predictor.
			(a) internal electric energy and b) scattering cross section of a \(200\,\)nm width nanorod as function of its length under TE polarized illumination (\(\mathbf{E}_{\text{in}}\) along the nanorod width).
			(c) averaged, \(XY\) projected internal field intensity and electric field vectors (real part, red arrows) at the anapole-type state in a Si block of dimensions \(240\times 220 \times 200\,\)nm\(^3\).
			(d-e) Contributions to the extinction coefficient of the electric dipole (d) and the magnetic dipole moment (e) in a silicon nanorod as function of its length. Top and bottom panels: nanorods excited under TE (\(\mathbf{E}\) along width) and TM (\(\mathbf{E}\) along length) polarizations, respectively. Left and right columns: network prediction and according numerical simulation, which are pairwise normalized to identical color scales. 3D illustrations of the internal electric field (real part) are shown for ANN and simulations at selected parameters.
			(f) internal fields and scattering radiation pattern for (i) Kerker-type forward scattering due to simultaneous excitation of similarly strong electric dipole (ED) and magnetic dipole (MD) modes for a cuboid of side length \(L=100\,\)nm (height \(H=200\,\)nm). (ii) Same for a larger cube, the MD dominates the optical response, hence bi-directional scattering is observed (\(L=160\,\)nm, \(H=200\,\)nm).
			(g) Forward/backward resolved scattering for a nanorod as function of its length under (i) TE and (ii) TM polarized plane wave illumination. Top: ANN predicted and simulated internal electric field shown for the forward scattering condition at the magnetic quadrupole mode.
		}
		\label{fig:derived_effects}
	\end{figure*}
	
	While a more formal benchmarking of the ANN is discussed further below, we start our results by demonstrating the ability of the network to capture some of the well-known physical effects in metallic and dielectric nanostructures. The ANN's capability to generalize to these arbitrary situations is tested by constructing a number of specific cases at which optical effects occur and compare the predictions of the neural network to numerical simulations. Figure~\ref{fig:planar_plasmonics}a shows the ANN predicted real and imaginary part of the amplitude of the internal field components \(E_x\) and \(E_y\) upon \(X\) polarized plane wave illumination of the planar gold nano-polygon.\cite{viarbitskaya_tailoring_2013, frank_short-range_2017}
	Clearly, the neural network correctly predicts the form and distribution of fields as given by the CDA simulation. In addition to arbitrary objects, the ANN shows a good systematic scaling of antenna behaviour, as illustrated in Figure~\ref{fig:planar_plasmonics}b where we plot the magnitude of the electric field intensity for gold nanorod antennas of \(75\,\)nm width and lengths ranging from a point dipole, to half-wave and multi-order antenna resonances. 
	In figure~\ref{fig:planar_plasmonics}c, the electric field intensity on top of the center of the nanorod is shown as function of the rod length.
	The neural network is able to make an accurate prediction of the mode structure and effective scaling of plasmonic nanoantennas within the domain under study in a generalized, phenomenological model of light-matter interaction.\cite{novotny_effective_2007}
	We note that the peak field enhancement at resonance is consistently underestimated by the ANN. We believe this is due to the random generation process of the training data, which therefore includes only very few structures that exactly hit a resonance condition.
	
	In Figure~\ref{fig:planar_plasmonics}d we finally show the scattering cross section (SCS) of a gold rod antenna as function of its length for perpendicular incident polarizations TE (incident field polarized perpendicular to the rod long axis) and TM (field along the rod axis).
	Again the neural network reliably predicts the resonance positions.
	Interestingly, apart from the dipole antenna resonance where the underestimated nearfield is reflected also in the SCS, prediction and scattering quantitatively matches better for longer rod antennas. This is a result of the weak coupling to the far-field of higher order modes, hence the underestimation of the nearfield strength has no great impact on the far-field scattering in these cases.

	Next to plasmonic nanoparticles and antennas, recently dielectric nanostructures have received tremendous interest for use in meta-surfaces and low-absorption antennas.\cite{kuznetsov_optically_2016, barredaRecentAdvancesHigh2019}
	Here, we evaluate the performance of the ANN in inferring the response of silicon nanorods with fixed height of 200~nm and varying lengths and widths.
	Figure~\ref{fig:derived_effects}a and b shows the internal electric field energy (a) and scattering cross section (b) for a silicon nanorod with a square cross-section of $200 \times 200$~nm$^2$, plotted against rod length. The incident plane wave is linearly polarized along the width of the rod (transverse electric, TE). In this dataset we observe the typical signature of an ``anapole'' mode for a length indicated by a dashed vertical line, at which a superposition of toroidal and electric dipole mode is excited.\cite{miroshnichenko_nonradiating_2015} The anapole mode is characterized by a minimum in the scattering cross section accompanied by a maximum of the electric field energy inside the structure.\cite{grinblat_enhanced_2016} Both signatures are reproduced by the ANN (solid lines) in excellent agreement with the numerical simulations (dashed lines). The corresponding field profile predicted by the ANN is shown in Figure~\ref{fig:derived_effects}c and reproduces the characteristic internal field distribution of the anapole. 
	
	The response of the silicon nanorods for polarizations along the width (TE) and along the length (TM) of the nanorod is further investigated in a parameter study where we independently tuned the length and width of the structure. Figure \ref{fig:derived_effects}d and \ref{fig:derived_effects}e show the electric and magnetic dipole contributions to the nanorod extinctions. The two separate contributions were extracted from the total solution through a Taylor-like expansion of the electric polarization.\cite{evlyukhin_multipole_2011} The neural network predictions (left subplots) agree very well with the numerical simulations (right subplots) in both the positions and amplitudes of the different magneto-electric modes in the structure. Also shown are 3D quiver plots of the internal field distributions at selected parameters, which show that the underlying field distributions inferred by the network match very well the simulated ones.

	\begin{figure*}[tb]
		\centering
		\includegraphics[page=1]{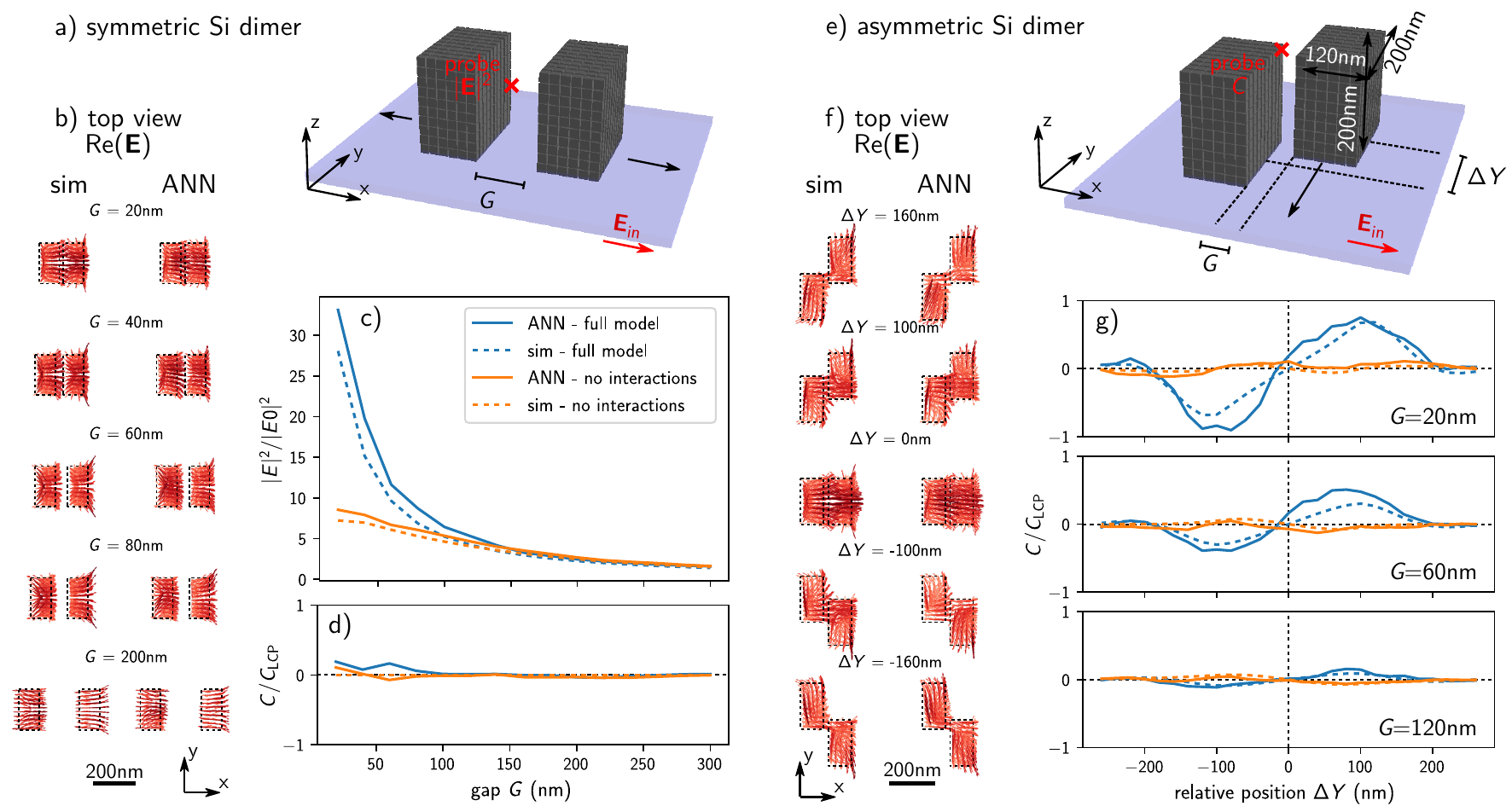}
		\caption{\FIGCAPTIONPREFIX
			Reproducibility of near-field enhancement and optical chirality of a silicon dimer.
			(a) Sketch of a symmetric dimer built of two identical silicon cuboids (\(120\times 200\times 200\,\)nm\(^3\)).
			(b) Top view of the electric field vectors inside the nanostructure upon \(X\) polarized plane wave illumination from above for different gaps sizes. Left column: numerical simulation, right column: ANN prediction.
			(c) Field enhancement in the center of the gap (red cross in (a)), calculated from the ANN prediction (solid lines) or with numerical simulations (dashed lines).
			(d) optical chirality \(C\) at \(30\,\)nm above the gap center (red cross in (e)), normalized to the chirality \(C_{\text{LCP}}\) of a left-circular polarized plane wave.
			(e)~Sketch of asymmetric silicon dimer. One constituent is vertically shifted by a distance \(\Delta Y\) relative to the other.
			(f)~Same as (b) for the asymmetric dimer and different relative positions \(\Delta Y\).
			(g) optical chirality \(C\), normalized to \(C_{\text{LCP}}\), calculated \(30\,\)nm above the middle of the gap, vertically centered at the left silicon block (red cross in (e)).
			Full field simulations (blue lines) are compared to calculations in which optical interactions have been artificially turned off (orange lines), see text.
			All data were taken for a normally incident plane wave (\(\mathbf{k}\) along \(-Z\), \(\lambda_0=700\,\)nm), with linear polarization along \(OX\).
			A video featuring the animated oscillating fields in the nano-dimer obtained from ANN and numerical simulations is available in the supporting material.
		}
		\label{fig:interactions_nearfield_chirality}
	\end{figure*}

	Precise tuning of the nanostructure geometry allows to obtain specific conditions of uni-directional ``Kerker-type'' light scattering due to the simultaneous excitation of electric dipole (ED) and magnetic dipole (MD) modes of similar magnitude.\cite{staude_tailoring_2013, wiecha_strongly_2017, shibanumaUnidirectionalLightScattering2016} For the fixed structure height of 200~nm and illumination wavelength used in our predictor network ($\lambda_0=700$~nm), we find that ED and MD modes are simultaneously excited for a square cuboid with side length \(L=100\,\)nm. Figure~\ref{fig:derived_effects}f shows a 3D quiver plot (i) where the MD can be observed in the upper part of the 3D electric field distribution (vortex formed by the field vectors), whereas the ED is situated at the bottom of the silicon block (field vectors parallel to the substrate plane). The superposition of both contributions leads to a strongly directional scattering pattern (see center plot). Panel (ii) shows the effect when the block size is increased to \(L=160\,\)nm. Here the MD becomes the predominantly excited mode, leading to a bi-directional scattering, as expected for a dipolar source of radiation. Both internal 3D field plots as well as the far-field scattering patterns show very good agreement between ANN and simulation.
	Generally the ratio of forward to backward total scattering is of interest in dielectric structures as it shows large variations with particle geometry. Figure~\ref{fig:derived_effects}g shows the forward / backward ratio for a silicon nanorod of \(100\,\)nm width and with increasing length for (i) TE and (ii) TM polarized illumination. The scattering from the nanorods was integrated over the respective full half hemisphere to obtain the total scattering intensity. Overall, the ANN predicts accurately the directional character of the scattering in all cases.
	The position of the quadrupole magnetic mode is indicated by the vertical line, even the quite complex electric field distributions at this position are correctly predicted by the network.
	For longer rods (\(\gtrsim 400\,\)nm), the accuracy of the ANN deteriorates slightly, the network tends to overestimate scattering and extinction cross sections (see Figure~\ref{fig:derived_effects}(d-e) and~\ref{fig:derived_effects}g). We assume that this is due to the maximum length and width of \(300\,\)nm used for the silicon blocks in the training data. 
	Considering this constraint on the training process, the network manages to generalize impressively well to larger individual nanostructures.
	In summary, our generalized predictor neural network manages to accurately predict various nano-optical phenomena in the near-field as well as in the far-field, without having been specifically trained on those effects. 
	
	Next to the response of individual structures, it is critically important that the ANN can capture the mutual interactions between isolated nanostructures. Figure~\ref{fig:interactions_nearfield_chirality} explores the near-field coupling for a silicon dimer structure. Results for gold nanodimers are shown in the Supporting Information Figure S6. The scaling of near-field coupling between the structures was studied by varying the inter-particle gap as illustrated in Figure~\ref{fig:interactions_nearfield_chirality}a. The effect of near-field coupling is clearly visible in the quiver plots in Figure~\ref{fig:interactions_nearfield_chirality}b. Visually the network predictions can barely be distinguished from the numerical simulations. Quantitative agreement is also obtained for the scaling of the near-field in the center of the gap (indicated by a red cross in Figure~\ref{fig:interactions_nearfield_chirality}a). The field enhancement derived via ANN is compared in Figure~\ref{fig:interactions_nearfield_chirality}c (solid lines) to numerical simulations (dashed lines), yielding indeed an excellent agreement, apart from a slight overestimation of the field intensity.
	
	In order to assess the impact of optical coupling between the two silicon blocks, we artificially turn off near-field interactions in the simulations (orange lines). This is done by calculating the optical response of both blocks separately, and using the internal dipole moments of those isolated simulations (respectively isolated ANN predictions) to derive the field intensity in the gap.	We find that near-field interactions become non-negligible only for small distances \(\lesssim 100\,\)nm, which is correctly described also by the predictor network. This observation is in agreement with experimental results of cathodoluminescence imaging of hybridized mode field profiles in silicon dimers.\cite{groep_direct_2016}
	
	In figure~\ref{fig:interactions_nearfield_chirality}d, we furthermore study the optical chirality \(C\) of the near-field in the center of the gap, \(10\,\)nm above the nanostructure's top surface. \(C\) is a measure of the selectivity at which opposite handed chiral molecules interact with the electromagnetic field.\cite{tang_optical_2010, meinzer_probing_2013} We plot \(C\) normalized to the chirality of a left circular polarized plane wave \(C_{\text{LCP}}\).  As expected, in the symmetric configuration the field has no chirality. It is however possible to break the symmetry of the dimer by shifting one of the blocks vertically by a distance \(\Delta Y\), as illustrated in Figure~\ref{fig:interactions_nearfield_chirality}e.
	The internal fields are shown in top-view quiver plots in Figure~\ref{fig:interactions_nearfield_chirality}f and provide evidence that the neural network correctly predicts near-field coupling effects in the asymmetric dimer. The electric field vectors inside the silicon strongly vary dependent on the relative positions of the two blocks. The optical chirality as function of the vertical shift is shown in Figure~\ref{fig:interactions_nearfield_chirality}g for three different gap sizes. Chirality is induced through asymmetry in the dimer geometry, which is again correctly reflected in the ANN predictions. Interestingly, the electromagnetic field above the gap features almost no chirality if the near-field interactions are deactivated in the simulations. Hence, the electromagnetic chirality is driven here by optical coupling between the two silicon blocks. This is confirmed by the decreasing magnitude of \(C\) for increasing gap widths (from top to bottom in Figure~\ref{fig:interactions_nearfield_chirality}g), which is also in agreement with recent experimental results.\cite{zhao_switchable_2019} In contrast, for plasmonic dimers, it is found that near-field coupling suppresses the chirality (Supporting Information Figure S6).\cite{meinzer_probing_2013}

	\begin{figure*}[tb]
		\centering
		\includegraphics[page=1]{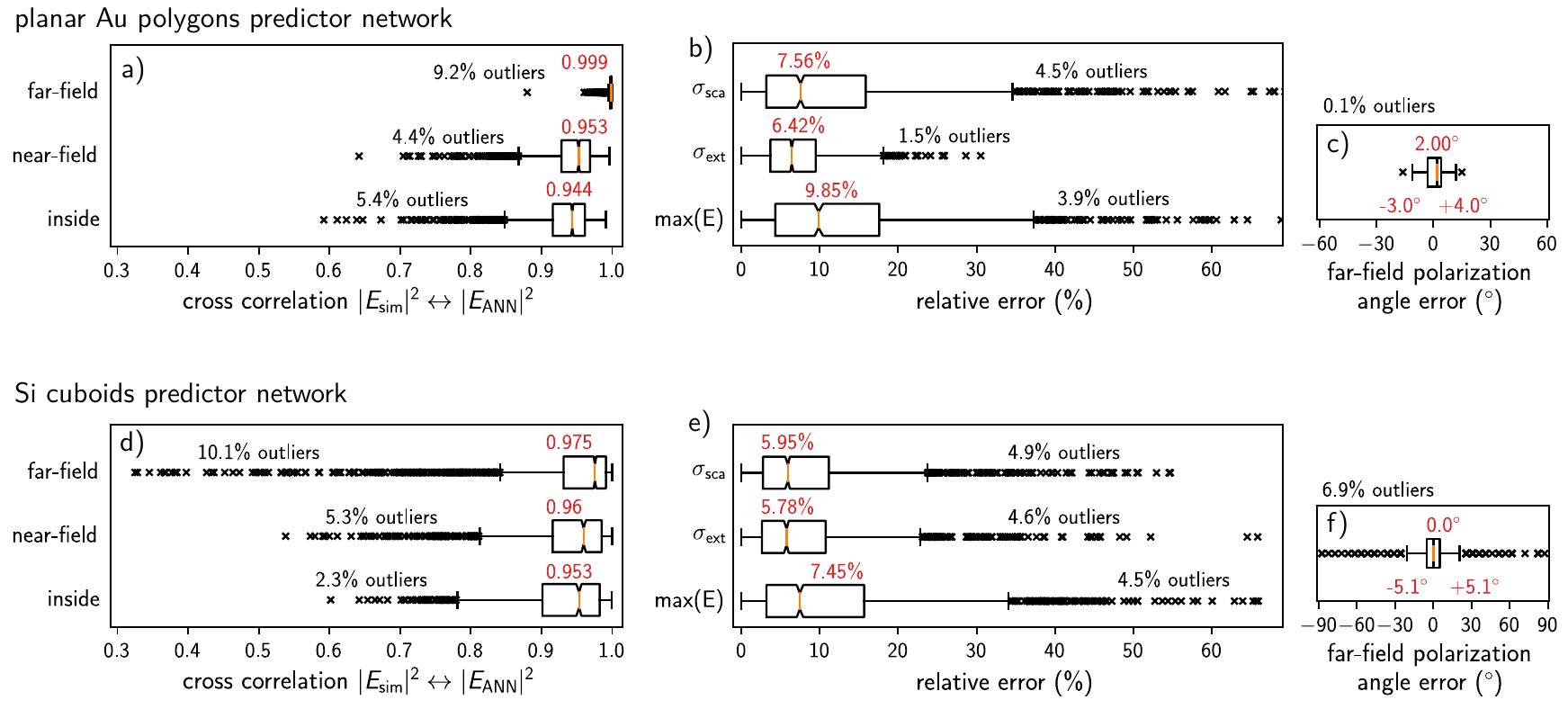}
		\caption{\FIGCAPTIONPREFIX
			Statistics on comparison between the ANN predictions and GDM simulations for the planar gold model (a-c) and the silicon model (d-f) on the validation data (not used for training). 
			(a,d) Qualitative comparison of field intensity distribution, in the far-field (comparing the back focal plane mappings), in the near-field in a plane \(45\,\)nm (a), respectively \(100\,\) (d) above the structure top surface, and inside the structure. 
			(b,e) Relative error of the ANN (in \%) for the scattering cross section, the extinction cross section, and the maximum value of near-field intensity. 
			(c,f) Absolute deviation in the far-field polarization angle. 
			All data points outside of 1.5 times the interquartile range are considered outliers and are individually plotted as cross symbols. Red numbers indicate the median values, and in case of (c,f) additionally the lower and upper quartiles.
			Specific examples of nanostructures from the validation sed  and the calculated physical observables are shown in the SI section~IV.
		}
		\label{fig:boxplots}
	\end{figure*}
	
	The above results reveal the capacity of the neural network in inferring many of the important physical effects in the interaction of light with nanostructures. While the simple model examples provide insight, a more formal assessment of the ANN performance is needed. Figure~\ref{fig:boxplots} shows results of the statistical data analysis of the entire validation set for the plasmonic (a-c) and silicon (d-f) models, each consisting of 2000 random structures which were not used for training. In Figure~\ref{fig:boxplots}a and~d the statistics for the normalized cross correlation between simulated and predicted field intensity distributions are shown for the far-field, the near-field in the vicinity, and the internal field inside the structure. Considering that a cross correlation of~0.8 indicates already a good qualitative agreement, the network predictions are mostly excellent, however for a non-negligible number of ``outliers'', the ANN results are significantly worse. 
	
	The same trends can be found in a quantitative analysis of the far-field cross-sections, the peak internal field intensity (Figure~\ref{fig:boxplots}b,e) and the polarization state (Figure~\ref{fig:boxplots}c,f). So while the network predictions are mostly very good, being a data-driven approach, an inevitable error of a few percent is inherent to the method. Also, the ANN bears some risk of getting a result with significant error. The risk of a failed prediction is of the order of around \(5\%\), as expressed by the amount of outliers in figure~\ref{fig:boxplots}. Examples of outliers in the validation data are shown in Figure~S4 and~S5 of the Supporting Information. In our simple examples we have also seen some reduction in ANN performance around resonant poles in the polarizability and for larger sized structures in Figure~\ref{fig:planar_plasmonics}. 
	The statistical evaluation indicates that the plasmonic predictor network performs generally worse than the ANN trained on silicon cuboidal structures. 
	The induced plasmonic currents cause strong depolarization effects on short length-scales, which are strongly dependent on the structure geometry. We therefore speculate that this complex and feature-rich plasmonic optical response is more difficult to predict, compared to dielectric nanostructures.
	This lack of perfection is the general property of neural networks and new methods gauging the reliability of solutions have been proposed using for example multiple, independently trained ANNs.\cite{wang_massive_2019} 
	Additionally, while ANNs are good at interpolating complex functions from few data, their performance is reduced when extrapolating outside of the known parameter space.\cite{nadell_deep_2019} 
	Despite these limitations, our results clearly show that the ANN's capabilities go far beyond that of a simple fit function, as it is able to apply trained behavior to new highly nontrivial configurations (see also Supporting Information Figure S8 on the prediction of curved structures). The ANN yields a generalized and powerful capability for generating field distributions from structural arrangements.
	
	Improved performance of the ANN in specific ranges may be achieved by further extending the training dataset. In our current demonstration we chose to limit ourselves to planar configurations and single materials. 
	However the method could be further generalized to arbitrary hybrid-material structures using the three-dimensional distribution of the dielectric constant as input. Periodic structures can be treated using training data describing one unit-cell of the periodic structure.\cite{gallinetElectromagneticScatteringFinite2009, evlyukhinOpticalResponseFeatures2010}
	In addition, the approach can be extended to spectrally resolved predictions using multiple output layers. 
	Spectral training might be accelerated by transfer learning. Generalized predictor networks can be potentially used together with evolutionary optimization schemes for fast, universal nano-photonic inverse design, overcoming the need of designing a specific inverse network for every target problem and nanostructure model.\cite{hegde_photonics_2020}
	
	In conclusion, we presented an approach combining the coupled dipole approximation with deep artificial neural networks, which is capable to accelerate universal electro-dynamical simulations of arbitrary 3D nanostructures by many orders of magnitude.
	Our generalized predictor network needs to be trained only once using a volume discretization of the three-dimensional electric polarization inside nanostructures of arbitrary shape. Being a data-driven approach, the method comes necessarily with a loss of accuracy. Despite this inherent shortcoming, we could demonstrate that the network predictions imply only a small error in the order of around five percent.
	On the other hand, the prediction of the internal fields allows to derive manifold quantities and effects in the near-field as well as in the far-field region.
	We have shown, that both regions are covered by the predictor with very good accuracy.
	We demonstrated furthermore that the network developed a generalized intuition about Maxwell's equations, being capable to reproduce complex nano-optical effects occurring in plasmonic and dielectric nanostructures. The ANN faithfully reproduces phenomena like higher order localized plasmon modes, magnetic and electric dipole modes, non-radiating anapole states, Kerker-type directional scattering or optical chirality. 
	We demonstrated that the network also developed an understanding of near-field interactions between separated nanostructures. We foresee that ultra-rapid generalized predictor networks bare tremendous potential for applications in nanophotonics.

	\begin{acknowledgments}
		We gratefully thank Arnaud Arbouet and Christian Girard for fruitful discussions. We thank the NVIDIA Corporation for the donation of a Quadro P6000 GPU used for this research.
		This work was supported by the German Research Foundation (DFG) through a research fellowship (WI 5261/1-1). OM acknowledges support through EPSRC grant EP/M009122/1.
		All data supporting this study are openly available from the University of Southampton repository (DOI: 10.5258/SOTON/D1088).
	\end{acknowledgments}

	\section*{Supporting Informations}
	A description of methods and further data with Figures S1-S10. 
	A video of the oscillating fields inside the silicon nano-dimer for varying vertical distances comparing ANN and simulation. 
	This material is available free of charge via the Internet at http://pubs.acs.org.

	\bibliography{2019_wiecha_generalized_predictor_network.bbl}

\providecommand{\latin}[1]{#1}
\makeatletter
\providecommand{\doi}
  {\begingroup\let\do\@makeother\dospecials
  \catcode`\{=1 \catcode`\}=2 \doi@aux}
\providecommand{\doi@aux}[1]{\endgroup\texttt{#1}}
\makeatother
\providecommand*\mcitethebibliography{\thebibliography}
\csname @ifundefined\endcsname{endmcitethebibliography}
  {\let\endmcitethebibliography\endthebibliography}{}
\begin{mcitethebibliography}{71}
\providecommand*\natexlab[1]{#1}
\providecommand*\mciteSetBstSublistMode[1]{}
\providecommand*\mciteSetBstMaxWidthForm[2]{}
\providecommand*\mciteBstWouldAddEndPuncttrue
  {\def\EndOfBibitem{\unskip.}}
\providecommand*\mciteBstWouldAddEndPunctfalse
  {\let\EndOfBibitem\relax}
\providecommand*\mciteSetBstMidEndSepPunct[3]{}
\providecommand*\mciteSetBstSublistLabelBeginEnd[3]{}
\providecommand*\EndOfBibitem{}
\mciteSetBstSublistMode{f}
\mciteSetBstMaxWidthForm{subitem}{(\alph{mcitesubitemcount})}
\mciteSetBstSublistLabelBeginEnd
  {\mcitemaxwidthsubitemform\space}
  {\relax}
  {\relax}

\bibitem[Draine and Flatau(1994)Draine, and
  Flatau]{draine_discrete-dipole_1994}
Draine,~B.~T.; Flatau,~P.~J. Discrete-Dipole Approximation for Scattering
  Calculations. \emph{Journal of the Optical Society of America A}
  \textbf{1994}, \emph{11}, 1491\relax
\mciteBstWouldAddEndPuncttrue
\mciteSetBstMidEndSepPunct{\mcitedefaultmidpunct}
{\mcitedefaultendpunct}{\mcitedefaultseppunct}\relax
\EndOfBibitem
\bibitem[Evlyukhin \latin{et~al.}(2011)Evlyukhin, Reinhardt, and
  Chichkov]{evlyukhin_multipole_2011}
Evlyukhin,~A.~B.; Reinhardt,~C.; Chichkov,~B.~N. Multipole Light Scattering by
  Nonspherical Nanoparticles in the Discrete Dipole Approximation.
  \emph{Physical Review B} \textbf{2011}, \emph{84}, 235429\relax
\mciteBstWouldAddEndPuncttrue
\mciteSetBstMidEndSepPunct{\mcitedefaultmidpunct}
{\mcitedefaultendpunct}{\mcitedefaultseppunct}\relax
\EndOfBibitem
\bibitem[Kinkhabwala \latin{et~al.}(2009)Kinkhabwala, Yu, Fan, Avlasevich,
  M{\"u}llen, and Moerner]{kinkhabwala_large_2009}
Kinkhabwala,~A.; Yu,~Z.; Fan,~S.; Avlasevich,~Y.; M{\"u}llen,~K.;
  Moerner,~W.~E. Large Single-Molecule Fluorescence Enhancements Produced by a
  Bowtie Nanoantenna. \emph{Nature Photonics} \textbf{2009}, \emph{3},
  654--657\relax
\mciteBstWouldAddEndPuncttrue
\mciteSetBstMidEndSepPunct{\mcitedefaultmidpunct}
{\mcitedefaultendpunct}{\mcitedefaultseppunct}\relax
\EndOfBibitem
\bibitem[Albella \latin{et~al.}(2013)Albella, Poyli, Schmidt, Maier, Moreno,
  S{\'a}enz, and Aizpurua]{albellaLowLossElectricMagnetic2013}
Albella,~P.; Poyli,~M.~A.; Schmidt,~M.~K.; Maier,~S.~A.; Moreno,~F.;
  S{\'a}enz,~J.~J.; Aizpurua,~J. Low-{{Loss Electric}} and {{Magnetic
  Field}}-{{Enhanced Spectroscopy}} with {{Subwavelength Silicon Dimers}}.
  \emph{The Journal of Physical Chemistry C} \textbf{2013}, \emph{117},
  13573--13584\relax
\mciteBstWouldAddEndPuncttrue
\mciteSetBstMidEndSepPunct{\mcitedefaultmidpunct}
{\mcitedefaultendpunct}{\mcitedefaultseppunct}\relax
\EndOfBibitem
\bibitem[Wiecha \latin{et~al.}(2017)Wiecha, Black, Wang, Paillard, Girard,
  Muskens, and Arbouet]{wiecha_polarization_2017}
Wiecha,~P.~R.; Black,~L.-J.; Wang,~Y.; Paillard,~V.; Girard,~C.;
  Muskens,~O.~L.; Arbouet,~A. Polarization Conversion in Plasmonic Nanoantennas
  for Metasurfaces Using Structural Asymmetry and Mode Hybridization.
  \emph{Scientific Reports} \textbf{2017}, \emph{7}, 40906\relax
\mciteBstWouldAddEndPuncttrue
\mciteSetBstMidEndSepPunct{\mcitedefaultmidpunct}
{\mcitedefaultendpunct}{\mcitedefaultseppunct}\relax
\EndOfBibitem
\bibitem[Sch{\"a}ferling \latin{et~al.}(2012)Sch{\"a}ferling, Dregely,
  Hentschel, and Giessen]{schaferling_tailoring_2012}
Sch{\"a}ferling,~M.; Dregely,~D.; Hentschel,~M.; Giessen,~H. Tailoring
  {{Enhanced Optical Chirality}}: {{Design Principles}} for {{Chiral Plasmonic
  Nanostructures}}. \emph{Physical Review X} \textbf{2012}, \emph{2},
  031010\relax
\mciteBstWouldAddEndPuncttrue
\mciteSetBstMidEndSepPunct{\mcitedefaultmidpunct}
{\mcitedefaultendpunct}{\mcitedefaultseppunct}\relax
\EndOfBibitem
\bibitem[Baffou \latin{et~al.}(2010)Baffou, Quidant, and
  Girard]{baffou_thermoplasmonics_2010}
Baffou,~G.; Quidant,~R.; Girard,~C. Thermoplasmonics Modeling: {{A Green}}'s
  Function Approach. \emph{Physical Review B} \textbf{2010}, \emph{82},
  165424\relax
\mciteBstWouldAddEndPuncttrue
\mciteSetBstMidEndSepPunct{\mcitedefaultmidpunct}
{\mcitedefaultendpunct}{\mcitedefaultseppunct}\relax
\EndOfBibitem
\bibitem[Kauranen and Zayats(2012)Kauranen, and
  Zayats]{kauranen_nonlinear_2012}
Kauranen,~M.; Zayats,~A.~V. Nonlinear Plasmonics. \emph{Nature Photonics}
  \textbf{2012}, \emph{6}, 737--748\relax
\mciteBstWouldAddEndPuncttrue
\mciteSetBstMidEndSepPunct{\mcitedefaultmidpunct}
{\mcitedefaultendpunct}{\mcitedefaultseppunct}\relax
\EndOfBibitem
\bibitem[Shcherbakov \latin{et~al.}(2014)Shcherbakov, Neshev, Hopkins,
  Shorokhov, Staude, {Melik-Gaykazyan}, Decker, Ezhov, Miroshnichenko, Brener,
  Fedyanin, and Kivshar]{shcherbakov_enhanced_2014}
Shcherbakov,~M.~R.; Neshev,~D.~N.; Hopkins,~B.; Shorokhov,~A.~S.; Staude,~I.;
  {Melik-Gaykazyan},~E.~V.; Decker,~M.; Ezhov,~A.~A.; Miroshnichenko,~A.~E.;
  Brener,~I.; Fedyanin,~A.~A.; Kivshar,~Y.~S. Enhanced {{Third}}-{{Harmonic
  Generation}} in {{Silicon Nanoparticles Driven}} by {{Magnetic Response}}.
  \emph{Nano Letters} \textbf{2014}, \emph{14}, 6488--6492\relax
\mciteBstWouldAddEndPuncttrue
\mciteSetBstMidEndSepPunct{\mcitedefaultmidpunct}
{\mcitedefaultendpunct}{\mcitedefaultseppunct}\relax
\EndOfBibitem
\bibitem[Wiecha \latin{et~al.}(2016)Wiecha, Arbouet, Girard, Baron, and
  Paillard]{wiecha_origin_2016}
Wiecha,~P.~R.; Arbouet,~A.; Girard,~C.; Baron,~T.; Paillard,~V. Origin of
  Second-Harmonic Generation from Individual Silicon Nanowires. \emph{Physical
  Review B} \textbf{2016}, \emph{93}, 125421\relax
\mciteBstWouldAddEndPuncttrue
\mciteSetBstMidEndSepPunct{\mcitedefaultmidpunct}
{\mcitedefaultendpunct}{\mcitedefaultseppunct}\relax
\EndOfBibitem
\bibitem[Maxwell(1865)]{maxwell_dynamical_1865}
Maxwell,~J.~C. A {{Dynamical Theory}} of the {{Electromagnetic Field}}.
  \emph{Philosophical Transactions of the Royal Society of London}
  \textbf{1865}, \emph{155}, 459--512\relax
\mciteBstWouldAddEndPuncttrue
\mciteSetBstMidEndSepPunct{\mcitedefaultmidpunct}
{\mcitedefaultendpunct}{\mcitedefaultseppunct}\relax
\EndOfBibitem
\bibitem[Smajic \latin{et~al.}(2009)Smajic, Hafner, Raguin, Tavzarashvili, and
  Mishrikey]{smajic_comparison_2009}
Smajic,~J.; Hafner,~C.; Raguin,~L.; Tavzarashvili,~K.; Mishrikey,~M. Comparison
  of {{Numerical Methods}} for the {{Analysis}} of {{Plasmonic Structures}}.
  \emph{Journal of Computational and Theoretical Nanoscience} \textbf{2009},
  \emph{6}, 763--774\relax
\mciteBstWouldAddEndPuncttrue
\mciteSetBstMidEndSepPunct{\mcitedefaultmidpunct}
{\mcitedefaultendpunct}{\mcitedefaultseppunct}\relax
\EndOfBibitem
\bibitem[Liu \latin{et~al.}(2018)Liu, Tan, Khoram, and Yu]{liu_training_2018}
Liu,~D.; Tan,~Y.; Khoram,~E.; Yu,~Z. Training {{Deep Neural Networks}} for the
  {{Inverse Design}} of {{Nanophotonic Structures}}. \emph{ACS Photonics}
  \textbf{2018}, \emph{5}, 1365--1369\relax
\mciteBstWouldAddEndPuncttrue
\mciteSetBstMidEndSepPunct{\mcitedefaultmidpunct}
{\mcitedefaultendpunct}{\mcitedefaultseppunct}\relax
\EndOfBibitem
\bibitem[Feichtner \latin{et~al.}(2012)Feichtner, Selig, Kiunke, and
  Hecht]{feichtner_evolutionary_2012}
Feichtner,~T.; Selig,~O.; Kiunke,~M.; Hecht,~B. Evolutionary {{Optimization}}
  of {{Optical Antennas}}. \emph{Physical Review Letters} \textbf{2012},
  \emph{109}, 127701\relax
\mciteBstWouldAddEndPuncttrue
\mciteSetBstMidEndSepPunct{\mcitedefaultmidpunct}
{\mcitedefaultendpunct}{\mcitedefaultseppunct}\relax
\EndOfBibitem
\bibitem[Wiecha \latin{et~al.}(2017)Wiecha, Arbouet, Girard, Lecestre, Larrieu,
  and Paillard]{wiecha_evolutionary_2017}
Wiecha,~P.~R.; Arbouet,~A.; Girard,~C.; Lecestre,~A.; Larrieu,~G.; Paillard,~V.
  Evolutionary Multi-Objective Optimization of Colour Pixels Based on
  Dielectric Nanoantennas. \emph{Nature Nanotechnology} \textbf{2017},
  \emph{12}, 163--169\relax
\mciteBstWouldAddEndPuncttrue
\mciteSetBstMidEndSepPunct{\mcitedefaultmidpunct}
{\mcitedefaultendpunct}{\mcitedefaultseppunct}\relax
\EndOfBibitem
\bibitem[Kingston \latin{et~al.}(2019)Kingston, Syed, Ngai, Sindhwani, and
  Chan]{kingston_assessing_2019}
Kingston,~B.~R.; Syed,~A.~M.; Ngai,~J.; Sindhwani,~S.; Chan,~W. C.~W. Assessing
  Micrometastases as a Target for Nanoparticles Using {{3D}} Microscopy and
  Machine Learning. \emph{Proceedings of the National Academy of Sciences}
  \textbf{2019}, 201907646\relax
\mciteBstWouldAddEndPuncttrue
\mciteSetBstMidEndSepPunct{\mcitedefaultmidpunct}
{\mcitedefaultendpunct}{\mcitedefaultseppunct}\relax
\EndOfBibitem
\bibitem[Liu \latin{et~al.}(2011)Liu, Hentschel, Weiss, Alivisatos, and
  Giessen]{liu_three-dimensional_2011}
Liu,~N.; Hentschel,~M.; Weiss,~T.; Alivisatos,~A.~P.; Giessen,~H.
  Three-{{Dimensional Plasmon Rulers}}. \emph{Science} \textbf{2011},
  \emph{332}, 1407--1410\relax
\mciteBstWouldAddEndPuncttrue
\mciteSetBstMidEndSepPunct{\mcitedefaultmidpunct}
{\mcitedefaultendpunct}{\mcitedefaultseppunct}\relax
\EndOfBibitem
\bibitem[Kuzyk \latin{et~al.}(2014)Kuzyk, Schreiber, Zhang, Govorov, Liedl, and
  Liu]{kuzyk_reconfigurable_2014}
Kuzyk,~A.; Schreiber,~R.; Zhang,~H.; Govorov,~A.~O.; Liedl,~T.; Liu,~N.
  Reconfigurable {{3D}} Plasmonic Metamolecules. \emph{Nature Materials}
  \textbf{2014}, \emph{13}, 862--866\relax
\mciteBstWouldAddEndPuncttrue
\mciteSetBstMidEndSepPunct{\mcitedefaultmidpunct}
{\mcitedefaultendpunct}{\mcitedefaultseppunct}\relax
\EndOfBibitem
\bibitem[Zhou \latin{et~al.}(2017)Zhou, Duan, and
  Liu]{zhou_dna-nanotechnology-enabled_2017}
Zhou,~C.; Duan,~X.; Liu,~N. {{DNA}}-{{Nanotechnology}}-{{Enabled Chiral
  Plasmonics}}: {{From Static}} to {{Dynamic}}. \emph{Accounts of Chemical
  Research} \textbf{2017}, \emph{50}, 2906--2914\relax
\mciteBstWouldAddEndPuncttrue
\mciteSetBstMidEndSepPunct{\mcitedefaultmidpunct}
{\mcitedefaultendpunct}{\mcitedefaultseppunct}\relax
\EndOfBibitem
\bibitem[Qian and Ginger(2017)Qian, and Ginger]{qian_reversibly_2017}
Qian,~Z.; Ginger,~D.~S. Reversibly {{Reconfigurable Colloidal Plasmonic
  Nanomaterials}}. \emph{Journal of the American Chemical Society}
  \textbf{2017}, \emph{139}, 5266--5276\relax
\mciteBstWouldAddEndPuncttrue
\mciteSetBstMidEndSepPunct{\mcitedefaultmidpunct}
{\mcitedefaultendpunct}{\mcitedefaultseppunct}\relax
\EndOfBibitem
\bibitem[Goodfellow \latin{et~al.}(2016)Goodfellow, Bengio, and
  Courville]{goodfellow_deep_2016}
Goodfellow,~I.; Bengio,~Y.; Courville,~A. \emph{Deep {{Learning}}}; {MIT
  Press}, 2016\relax
\mciteBstWouldAddEndPuncttrue
\mciteSetBstMidEndSepPunct{\mcitedefaultmidpunct}
{\mcitedefaultendpunct}{\mcitedefaultseppunct}\relax
\EndOfBibitem
\bibitem[LeCun \latin{et~al.}(2015)LeCun, Bengio, and Hinton]{lecun_deep_2015}
LeCun,~Y.; Bengio,~Y.; Hinton,~G. Deep Learning. \emph{Nature} \textbf{2015},
  \emph{521}, 436--444\relax
\mciteBstWouldAddEndPuncttrue
\mciteSetBstMidEndSepPunct{\mcitedefaultmidpunct}
{\mcitedefaultendpunct}{\mcitedefaultseppunct}\relax
\EndOfBibitem
\bibitem[Rivenson \latin{et~al.}(2017)Rivenson, Zhang, Gunaydin, Teng, and
  Ozcan]{rivenson_phase_2017}
Rivenson,~Y.; Zhang,~Y.; Gunaydin,~H.; Teng,~D.; Ozcan,~A. Phase Recovery and
  Holographic Image Reconstruction Using Deep Learning in Neural Networks.
  \emph{arXiv:1705.04286 [physics]} \textbf{2017}, \relax
\mciteBstWouldAddEndPunctfalse
\mciteSetBstMidEndSepPunct{\mcitedefaultmidpunct}
{}{\mcitedefaultseppunct}\relax
\EndOfBibitem
\bibitem[Baumeister \latin{et~al.}(2018)Baumeister, Brunton, and
  Kutz]{baumeister_deep_2018-1}
Baumeister,~T.; Brunton,~S.~L.; Kutz,~J.~N. Deep Learning and Model Predictive
  Control for Self-Tuning Mode-Locked Lasers. \emph{JOSA B} \textbf{2018},
  \emph{35}, 617--626\relax
\mciteBstWouldAddEndPuncttrue
\mciteSetBstMidEndSepPunct{\mcitedefaultmidpunct}
{\mcitedefaultendpunct}{\mcitedefaultseppunct}\relax
\EndOfBibitem
\bibitem[Cirovic(1997)]{cirovic_feed-forward_1997}
Cirovic,~D.~A. Feed-Forward Artificial Neural Networks: Applications to
  Spectroscopy. \emph{TrAC Trends in Analytical Chemistry} \textbf{1997},
  \emph{16}, 148--155\relax
\mciteBstWouldAddEndPuncttrue
\mciteSetBstMidEndSepPunct{\mcitedefaultmidpunct}
{\mcitedefaultendpunct}{\mcitedefaultseppunct}\relax
\EndOfBibitem
\bibitem[Ciresan \latin{et~al.}(2012)Ciresan, Giusti, Gambardella, and
  Schmidhuber]{ciresan_deep_2012}
Ciresan,~D.; Giusti,~A.; Gambardella,~L.~M.; Schmidhuber,~J. In \emph{Advances
  in {{Neural Information Processing Systems}} 25}; Pereira,~F., Burges,~C.
  J.~C., Bottou,~L., Weinberger,~K.~Q., Eds.; {Curran Associates, Inc.}, 2012;
  pp 2843--2851\relax
\mciteBstWouldAddEndPuncttrue
\mciteSetBstMidEndSepPunct{\mcitedefaultmidpunct}
{\mcitedefaultendpunct}{\mcitedefaultseppunct}\relax
\EndOfBibitem
\bibitem[Wang \latin{et~al.}(2014)Wang, Cancilla, Torrecilla, and
  Haick]{wang_artificial_2014}
Wang,~B.; Cancilla,~J.~C.; Torrecilla,~J.~S.; Haick,~H. Artificial {{Sensing
  Intelligence}} with {{Silicon Nanowires}} for {{Ultraselective Detection}} in
  the {{Gas Phase}}. \emph{Nano Letters} \textbf{2014}, \emph{14},
  933--938\relax
\mciteBstWouldAddEndPuncttrue
\mciteSetBstMidEndSepPunct{\mcitedefaultmidpunct}
{\mcitedefaultendpunct}{\mcitedefaultseppunct}\relax
\EndOfBibitem
\bibitem[Ziatdinov \latin{et~al.}(2017)Ziatdinov, Dyck, Maksov, Li, Sang, Xiao,
  Unocic, Vasudevan, Jesse, and Kalinin]{ziatdinov_deep_2017}
Ziatdinov,~M.; Dyck,~O.; Maksov,~A.; Li,~X.; Sang,~X.; Xiao,~K.; Unocic,~R.~R.;
  Vasudevan,~R.; Jesse,~S.; Kalinin,~S.~V. Deep {{Learning}} of {{Atomically
  Resolved Scanning Transmission Electron Microscopy Images}}: {{Chemical
  Identification}} and {{Tracking Local Transformations}}. \emph{ACS Nano}
  \textbf{2017}, \emph{11}, 12742--12752\relax
\mciteBstWouldAddEndPuncttrue
\mciteSetBstMidEndSepPunct{\mcitedefaultmidpunct}
{\mcitedefaultendpunct}{\mcitedefaultseppunct}\relax
\EndOfBibitem
\bibitem[Jo \latin{et~al.}(2017)Jo, Park, Jung, Yoon, Joo, Kim, Kang, Choi,
  Lee, and Park]{jo_holographic_2017}
Jo,~Y.; Park,~S.; Jung,~J.; Yoon,~J.; Joo,~H.; Kim,~M.-h.; Kang,~S.-J.;
  Choi,~M.~C.; Lee,~S.~Y.; Park,~Y. Holographic Deep Learning for Rapid Optical
  Screening of Anthrax Spores. \emph{Science Advances} \textbf{2017}, \emph{3},
  e1700606\relax
\mciteBstWouldAddEndPuncttrue
\mciteSetBstMidEndSepPunct{\mcitedefaultmidpunct}
{\mcitedefaultendpunct}{\mcitedefaultseppunct}\relax
\EndOfBibitem
\bibitem[Zhang \latin{et~al.}(2018)Zhang, Liu, Chaurasia, Ma, Mlodzianoski,
  Culurciello, and Huang]{zhang_analyzing_2018}
Zhang,~P.; Liu,~S.; Chaurasia,~A.; Ma,~D.; Mlodzianoski,~M.~J.;
  Culurciello,~E.; Huang,~F. Analyzing Complex Single-Molecule Emission
  Patterns with Deep Learning. \emph{Nature Methods} \textbf{2018}, \emph{15},
  913\relax
\mciteBstWouldAddEndPuncttrue
\mciteSetBstMidEndSepPunct{\mcitedefaultmidpunct}
{\mcitedefaultendpunct}{\mcitedefaultseppunct}\relax
\EndOfBibitem
\bibitem[Han \latin{et~al.}(2019)Han, Lin, Yang, Mao, Li, Wang, Fatemi, Zhou,
  Wang, Ma, Cao, {Rodan-Legrain}, Bie, {Navarro-Moratalla}, Klein, MacNeill,
  Wu, Leong, Kitadai, Ling, {Jarillo-Herrero}, Palacios, Yin, and
  Kong]{han_deep_2019}
Han,~B. \latin{et~al.}  Deep {{Learning Enabled Fast Optical Characterization}}
  of {{Two}}-{{Dimensional Materials}}. \emph{arXiv:1906.11220 [cond-mat,
  physics:physics]} \textbf{2019}, \relax
\mciteBstWouldAddEndPunctfalse
\mciteSetBstMidEndSepPunct{\mcitedefaultmidpunct}
{}{\mcitedefaultseppunct}\relax
\EndOfBibitem
\bibitem[Wiecha \latin{et~al.}(2019)Wiecha, Lecestre, Mallet, and
  Larrieu]{wiecha_pushing_2019}
Wiecha,~P.~R.; Lecestre,~A.; Mallet,~N.; Larrieu,~G. Pushing the Limits of
  Optical Information Storage Using Deep Learning. \emph{Nature Nanotechnology}
  \textbf{2019}, 1\relax
\mciteBstWouldAddEndPuncttrue
\mciteSetBstMidEndSepPunct{\mcitedefaultmidpunct}
{\mcitedefaultendpunct}{\mcitedefaultseppunct}\relax
\EndOfBibitem
\bibitem[Timoshenko \latin{et~al.}(2019)Timoshenko, Wrasman, Luneau, Shirman,
  Cargnello, Bare, Aizenberg, Friend, and Frenkel]{timoshenko_probing_2019}
Timoshenko,~J.; Wrasman,~C.~J.; Luneau,~M.; Shirman,~T.; Cargnello,~M.;
  Bare,~S.~R.; Aizenberg,~J.; Friend,~C.~M.; Frenkel,~A.~I. Probing {{Atomic
  Distributions}} in {{Mono}}- and {{Bimetallic Nanoparticles}} by {{Supervised
  Machine Learning}}. \emph{Nano Letters} \textbf{2019}, \emph{19},
  520--529\relax
\mciteBstWouldAddEndPuncttrue
\mciteSetBstMidEndSepPunct{\mcitedefaultmidpunct}
{\mcitedefaultendpunct}{\mcitedefaultseppunct}\relax
\EndOfBibitem
\bibitem[Selle \latin{et~al.}(2008)Selle, Brixner, Bayer, Wollenhaupt, and
  Baumert]{selle_modelling_2008}
Selle,~R.; Brixner,~T.; Bayer,~T.; Wollenhaupt,~M.; Baumert,~T. Modelling of
  Ultrafast Coherent Strong-Field Dynamics in Potassium with Neural Networks.
  \emph{Journal of Physics B: Atomic, Molecular and Optical Physics}
  \textbf{2008}, \emph{41}, 074019\relax
\mciteBstWouldAddEndPuncttrue
\mciteSetBstMidEndSepPunct{\mcitedefaultmidpunct}
{\mcitedefaultendpunct}{\mcitedefaultseppunct}\relax
\EndOfBibitem
\bibitem[Selle \latin{et~al.}(2007)Selle, Vogt, Brixner, Gerber, Metzler, and
  Kinzel]{selle_modeling_2007}
Selle,~R.; Vogt,~G.; Brixner,~T.; Gerber,~G.; Metzler,~R.; Kinzel,~W. Modeling
  of Light-Matter Interactions with Neural Networks. \emph{Physical Review A}
  \textbf{2007}, \emph{76}, 023810\relax
\mciteBstWouldAddEndPuncttrue
\mciteSetBstMidEndSepPunct{\mcitedefaultmidpunct}
{\mcitedefaultendpunct}{\mcitedefaultseppunct}\relax
\EndOfBibitem
\bibitem[Malkiel \latin{et~al.}(2018)Malkiel, Mrejen, Nagler, Arieli, Wolf, and
  Suchowski]{malkiel_plasmonic_2018}
Malkiel,~I.; Mrejen,~M.; Nagler,~A.; Arieli,~U.; Wolf,~L.; Suchowski,~H.
  Plasmonic Nanostructure Design and Characterization via {{Deep Learning}}.
  \emph{Light: Science \& Applications} \textbf{2018}, \emph{7}, 60\relax
\mciteBstWouldAddEndPuncttrue
\mciteSetBstMidEndSepPunct{\mcitedefaultmidpunct}
{\mcitedefaultendpunct}{\mcitedefaultseppunct}\relax
\EndOfBibitem
\bibitem[Peurifoy \latin{et~al.}(2018)Peurifoy, Shen, Jing, Yang,
  {Cano-Renteria}, DeLacy, Joannopoulos, Tegmark, and Solja{\v
  c}i{\'c}]{peurifoy_nanophotonic_2018}
Peurifoy,~J.; Shen,~Y.; Jing,~L.; Yang,~Y.; {Cano-Renteria},~F.; DeLacy,~B.~G.;
  Joannopoulos,~J.~D.; Tegmark,~M.; Solja{\v c}i{\'c},~M. Nanophotonic Particle
  Simulation and Inverse Design Using Artificial Neural Networks. \emph{Science
  Advances} \textbf{2018}, \emph{4}, eaar4206\relax
\mciteBstWouldAddEndPuncttrue
\mciteSetBstMidEndSepPunct{\mcitedefaultmidpunct}
{\mcitedefaultendpunct}{\mcitedefaultseppunct}\relax
\EndOfBibitem
\bibitem[Liu \latin{et~al.}(2018)Liu, Zhu, Rodrigues, Lee, and
  Cai]{liu_generative_2018}
Liu,~Z.; Zhu,~D.; Rodrigues,~S.~P.; Lee,~K.-T.; Cai,~W. Generative {{Model}}
  for the {{Inverse Design}} of {{Metasurfaces}}. \emph{Nano Letters}
  \textbf{2018}, \emph{18}, 6570--6576\relax
\mciteBstWouldAddEndPuncttrue
\mciteSetBstMidEndSepPunct{\mcitedefaultmidpunct}
{\mcitedefaultendpunct}{\mcitedefaultseppunct}\relax
\EndOfBibitem
\bibitem[Liu \latin{et~al.}(2019)Liu, Raju, Zhu, and Cai]{liu_hybrid_2019}
Liu,~Z.; Raju,~L.; Zhu,~D.; Cai,~W. A {{Hybrid Strategy}} for the {{Discovery}}
  and {{Design}} of {{Photonic Nanostructures}}. \emph{arXiv:1902.02293
  [physics]} \textbf{2019}, \relax
\mciteBstWouldAddEndPunctfalse
\mciteSetBstMidEndSepPunct{\mcitedefaultmidpunct}
{}{\mcitedefaultseppunct}\relax
\EndOfBibitem
\bibitem[An \latin{et~al.}(2019)An, Fowler, Zheng, Shalaginov, Tang, Li, Zhou,
  Ding, Agarwal, {Rivero-Baleine}, Richardson, Gu, Hu, and
  Zhang]{an_novel_2019}
An,~S.; Fowler,~C.; Zheng,~B.; Shalaginov,~M.~Y.; Tang,~H.; Li,~H.; Zhou,~L.;
  Ding,~J.; Agarwal,~A.~M.; {Rivero-Baleine},~C.; Richardson,~K.~A.; Gu,~T.;
  Hu,~J.; Zhang,~H. A {{Novel Modeling Approach}} for {{All}}-{{Dielectric
  Metasurfaces Using Deep Neural Networks}}. \emph{arXiv:1906.03387 [physics]}
  \textbf{2019}, \relax
\mciteBstWouldAddEndPunctfalse
\mciteSetBstMidEndSepPunct{\mcitedefaultmidpunct}
{}{\mcitedefaultseppunct}\relax
\EndOfBibitem
\bibitem[Chen \latin{et~al.}(2019)Chen, Zhu, Xie, Feng, and
  Liu]{chen_smart_2019}
Chen,~Y.; Zhu,~J.; Xie,~Y.; Feng,~N.; Liu,~Q. H.~H. Smart Inverse Design of
  Graphene-Based Photonic Metamaterials by an Adaptive Artificial Neural
  Network. \emph{Nanoscale} \textbf{2019}, \relax
\mciteBstWouldAddEndPunctfalse
\mciteSetBstMidEndSepPunct{\mcitedefaultmidpunct}
{}{\mcitedefaultseppunct}\relax
\EndOfBibitem
\bibitem[Liu \latin{et~al.}(2019)Liu, Zhu, Lee, Kim, Raju, and
  Cai]{liu_compounding_2019}
Liu,~Z.; Zhu,~D.; Lee,~K.-T.; Kim,~A.~S.; Raju,~L.; Cai,~W. Compounding
  Meta-Atoms into Meta-Molecules with Hybrid Artificial Intelligence
  Techniques. \emph{arXiv:1907.03366 [physics]} \textbf{2019}, \relax
\mciteBstWouldAddEndPunctfalse
\mciteSetBstMidEndSepPunct{\mcitedefaultmidpunct}
{}{\mcitedefaultseppunct}\relax
\EndOfBibitem
\bibitem[Jiang and Fan(2019)Jiang, and Fan]{jiang_global_2019}
Jiang,~J.; Fan,~J.~A. Global {{Optimization}} of {{Dielectric Metasurfaces
  Using}} a {{Physics}}-{{Driven Neural Network}}. \emph{Nano Letters}
  \textbf{2019}, \emph{19}, 5366--5372\relax
\mciteBstWouldAddEndPuncttrue
\mciteSetBstMidEndSepPunct{\mcitedefaultmidpunct}
{\mcitedefaultendpunct}{\mcitedefaultseppunct}\relax
\EndOfBibitem
\bibitem[Girard(2005)]{girard_near_2005}
Girard,~C. Near Fields in Nanostructures. \emph{Reports on Progress in Physics}
  \textbf{2005}, \emph{68}, 1883--1933\relax
\mciteBstWouldAddEndPuncttrue
\mciteSetBstMidEndSepPunct{\mcitedefaultmidpunct}
{\mcitedefaultendpunct}{\mcitedefaultseppunct}\relax
\EndOfBibitem
\bibitem[Wiecha(2018)]{wiecha_pygdmpython_2018}
Wiecha,~P.~R. {{pyGDM}}\textemdash{{A}} Python Toolkit for Full-Field
  Electro-Dynamical Simulations and Evolutionary Optimization of
  Nanostructures. \emph{Computer Physics Communications} \textbf{2018},
  \emph{233}, 167--192\relax
\mciteBstWouldAddEndPuncttrue
\mciteSetBstMidEndSepPunct{\mcitedefaultmidpunct}
{\mcitedefaultendpunct}{\mcitedefaultseppunct}\relax
\EndOfBibitem
\bibitem[Wiecha \latin{et~al.}(2017)Wiecha, Cuche, Arbouet, Girard, {Colas des
  Francs}, Lecestre, Larrieu, Fournel, Larrey, Baron, and
  Paillard]{wiecha_strongly_2017}
Wiecha,~P.~R.; Cuche,~A.; Arbouet,~A.; Girard,~C.; {Colas des Francs},~G.;
  Lecestre,~A.; Larrieu,~G.; Fournel,~F.; Larrey,~V.; Baron,~T.; Paillard,~V.
  Strongly {{Directional Scattering}} from {{Dielectric Nanowires}}. \emph{ACS
  Photonics} \textbf{2017}, \emph{4}, 2036--2046\relax
\mciteBstWouldAddEndPuncttrue
\mciteSetBstMidEndSepPunct{\mcitedefaultmidpunct}
{\mcitedefaultendpunct}{\mcitedefaultseppunct}\relax
\EndOfBibitem
\bibitem[Girard \latin{et~al.}(2018)Girard, Wiecha, Cuche, and
  Dujardin]{girard_designing_2018}
Girard,~C.; Wiecha,~P.~R.; Cuche,~A.; Dujardin,~E. Designing Thermoplasmonic
  Properties of Metallic Metasurfaces. \emph{Journal of Optics} \textbf{2018},
  \emph{20}, 075004\relax
\mciteBstWouldAddEndPuncttrue
\mciteSetBstMidEndSepPunct{\mcitedefaultmidpunct}
{\mcitedefaultendpunct}{\mcitedefaultseppunct}\relax
\EndOfBibitem
\bibitem[Balla \latin{et~al.}(2010)Balla, So, and Sheppard]{balla_second_2010}
Balla,~N.~K.; So,~P. T.~C.; Sheppard,~C. J.~R. Second Harmonic Scattering from
  Small Particles Using {{Discrete Dipole Approximation}}. \emph{Optics
  Express} \textbf{2010}, \emph{18}, 21603--21611\relax
\mciteBstWouldAddEndPuncttrue
\mciteSetBstMidEndSepPunct{\mcitedefaultmidpunct}
{\mcitedefaultendpunct}{\mcitedefaultseppunct}\relax
\EndOfBibitem
\bibitem[Teulle \latin{et~al.}(2012)Teulle, Marty, Viarbitskaya, Arbouet,
  Dujardin, Girard, and {Colas des Francs}]{teulle_scanning_2012}
Teulle,~A.; Marty,~R.; Viarbitskaya,~S.; Arbouet,~A.; Dujardin,~E.; Girard,~C.;
  {Colas des Francs},~G. Scanning Optical Microscopy Modeling in
  Nanoplasmonics. \emph{Journal of the Optical Society of America B}
  \textbf{2012}, \emph{29}, 2431\relax
\mciteBstWouldAddEndPuncttrue
\mciteSetBstMidEndSepPunct{\mcitedefaultmidpunct}
{\mcitedefaultendpunct}{\mcitedefaultseppunct}\relax
\EndOfBibitem
\bibitem[Ronneberger \latin{et~al.}(2015)Ronneberger, Fischer, and
  Brox]{ronneberger_u-net_2015}
Ronneberger,~O.; Fischer,~P.; Brox,~T. U-{{Net}}: {{Convolutional Networks}}
  for {{Biomedical Image Segmentation}}. \emph{arXiv:1505.04597 [cs]}
  \textbf{2015}, \relax
\mciteBstWouldAddEndPunctfalse
\mciteSetBstMidEndSepPunct{\mcitedefaultmidpunct}
{}{\mcitedefaultseppunct}\relax
\EndOfBibitem
\bibitem[He \latin{et~al.}(2015)He, Zhang, Ren, and Sun]{he_deep_2015}
He,~K.; Zhang,~X.; Ren,~S.; Sun,~J. Deep {{Residual Learning}} for {{Image
  Recognition}}. \emph{arXiv:1512.03385 [cs]} \textbf{2015}, \relax
\mciteBstWouldAddEndPunctfalse
\mciteSetBstMidEndSepPunct{\mcitedefaultmidpunct}
{}{\mcitedefaultseppunct}\relax
\EndOfBibitem
\bibitem[Szegedy \latin{et~al.}(2016)Szegedy, Ioffe, Vanhoucke, and
  Alemi]{szegedy_inception-v4_2016}
Szegedy,~C.; Ioffe,~S.; Vanhoucke,~V.; Alemi,~A. Inception-v4,
  {{Inception}}-{{ResNet}} and the {{Impact}} of {{Residual Connections}} on
  {{Learning}}. \emph{arXiv:1602.07261 [cs]} \textbf{2016}, \relax
\mciteBstWouldAddEndPunctfalse
\mciteSetBstMidEndSepPunct{\mcitedefaultmidpunct}
{}{\mcitedefaultseppunct}\relax
\EndOfBibitem
\bibitem[Ozcan and Sen(2012)Ozcan, and
  Sen]{ozcanInvestigationPerformanceLU2012}
Ozcan,~C.; Sen,~B. Investigation of the Performance of {{LU}} Decomposition
  Method Using {{CUDA}}. \emph{Procedia Technology} \textbf{2012}, \emph{1},
  50--54\relax
\mciteBstWouldAddEndPuncttrue
\mciteSetBstMidEndSepPunct{\mcitedefaultmidpunct}
{\mcitedefaultendpunct}{\mcitedefaultseppunct}\relax
\EndOfBibitem
\bibitem[Viarbitskaya \latin{et~al.}(2013)Viarbitskaya, Teulle, Marty, Sharma,
  Girard, Arbouet, and Dujardin]{viarbitskaya_tailoring_2013}
Viarbitskaya,~S.; Teulle,~A.; Marty,~R.; Sharma,~J.; Girard,~C.; Arbouet,~A.;
  Dujardin,~E. Tailoring and Imaging the Plasmonic Local Density of States in
  Crystalline Nanoprisms. \emph{Nature Materials} \textbf{2013}, \emph{12},
  426--432\relax
\mciteBstWouldAddEndPuncttrue
\mciteSetBstMidEndSepPunct{\mcitedefaultmidpunct}
{\mcitedefaultendpunct}{\mcitedefaultseppunct}\relax
\EndOfBibitem
\bibitem[Frank \latin{et~al.}(2017)Frank, Kahl, Podbiel, Spektor, Orenstein,
  Fu, Weiss, Hoegen, Davis, zu~Heringdorf, and Giessen]{frank_short-range_2017}
Frank,~B.; Kahl,~P.; Podbiel,~D.; Spektor,~G.; Orenstein,~M.; Fu,~L.;
  Weiss,~T.; Hoegen,~M. H.-v.; Davis,~T.~J.; zu~Heringdorf,~F.-J.~M.;
  Giessen,~H. Short-Range Surface Plasmonics: {{Localized}} Electron Emission
  Dynamics from a 60-Nm Spot on an Atomically Flat Single-Crystalline Gold
  Surface. \emph{Science Advances} \textbf{2017}, \emph{3}, e1700721\relax
\mciteBstWouldAddEndPuncttrue
\mciteSetBstMidEndSepPunct{\mcitedefaultmidpunct}
{\mcitedefaultendpunct}{\mcitedefaultseppunct}\relax
\EndOfBibitem
\bibitem[Novotny(2007)]{novotny_effective_2007}
Novotny,~L. Effective {{Wavelength Scaling}} for {{Optical Antennas}}.
  \emph{Physical Review Letters} \textbf{2007}, \emph{98}, 266802\relax
\mciteBstWouldAddEndPuncttrue
\mciteSetBstMidEndSepPunct{\mcitedefaultmidpunct}
{\mcitedefaultendpunct}{\mcitedefaultseppunct}\relax
\EndOfBibitem
\bibitem[Kuznetsov \latin{et~al.}(2016)Kuznetsov, Miroshnichenko, Brongersma,
  Kivshar, and Luk'yanchuk]{kuznetsov_optically_2016}
Kuznetsov,~A.~I.; Miroshnichenko,~A.~E.; Brongersma,~M.~L.; Kivshar,~Y.~S.;
  Luk'yanchuk,~B. Optically Resonant Dielectric Nanostructures. \emph{Science}
  \textbf{2016}, \emph{354}\relax
\mciteBstWouldAddEndPuncttrue
\mciteSetBstMidEndSepPunct{\mcitedefaultmidpunct}
{\mcitedefaultendpunct}{\mcitedefaultseppunct}\relax
\EndOfBibitem
\bibitem[Barreda \latin{et~al.}(2019)Barreda, Saiz, Gonz{\'a}lez, Moreno, and
  Albella]{barredaRecentAdvancesHigh2019}
Barreda,~A.~I.; Saiz,~J.~M.; Gonz{\'a}lez,~F.; Moreno,~F.; Albella,~P. Recent
  Advances in High Refractive Index Dielectric Nanoantennas: {{Basics}} and
  Applications. \emph{AIP Advances} \textbf{2019}, \emph{9}, 040701\relax
\mciteBstWouldAddEndPuncttrue
\mciteSetBstMidEndSepPunct{\mcitedefaultmidpunct}
{\mcitedefaultendpunct}{\mcitedefaultseppunct}\relax
\EndOfBibitem
\bibitem[Miroshnichenko \latin{et~al.}(2015)Miroshnichenko, Evlyukhin, Yu,
  Bakker, Chipouline, Kuznetsov, Luk'yanchuk, Chichkov, and
  Kivshar]{miroshnichenko_nonradiating_2015}
Miroshnichenko,~A.~E.; Evlyukhin,~A.~B.; Yu,~Y.~F.; Bakker,~R.~M.;
  Chipouline,~A.; Kuznetsov,~A.~I.; Luk'yanchuk,~B.; Chichkov,~B.~N.;
  Kivshar,~Y.~S. Nonradiating Anapole Modes in Dielectric Nanoparticles.
  \emph{Nature Communications} \textbf{2015}, \emph{6}, 8069\relax
\mciteBstWouldAddEndPuncttrue
\mciteSetBstMidEndSepPunct{\mcitedefaultmidpunct}
{\mcitedefaultendpunct}{\mcitedefaultseppunct}\relax
\EndOfBibitem
\bibitem[Grinblat \latin{et~al.}(2016)Grinblat, Li, Nielsen, Oulton, and
  Maier]{grinblat_enhanced_2016}
Grinblat,~G.; Li,~Y.; Nielsen,~M.~P.; Oulton,~R.~F.; Maier,~S.~A. Enhanced
  {{Third Harmonic Generation}} in {{Single Germanium Nanodisks Excited}} at
  the {{Anapole Mode}}. \emph{Nano Letters} \textbf{2016}, \relax
\mciteBstWouldAddEndPunctfalse
\mciteSetBstMidEndSepPunct{\mcitedefaultmidpunct}
{}{\mcitedefaultseppunct}\relax
\EndOfBibitem
\bibitem[Staude \latin{et~al.}(2013)Staude, Miroshnichenko, Decker, Fofang,
  Liu, Gonzales, Dominguez, Luk, Neshev, Brener, and
  Kivshar]{staude_tailoring_2013}
Staude,~I.; Miroshnichenko,~A.~E.; Decker,~M.; Fofang,~N.~T.; Liu,~S.;
  Gonzales,~E.; Dominguez,~J.; Luk,~T.~S.; Neshev,~D.~N.; Brener,~I.;
  Kivshar,~Y. Tailoring {{Directional Scattering}} through {{Magnetic}} and
  {{Electric Resonances}} in {{Subwavelength Silicon Nanodisks}}. \emph{ACS
  Nano} \textbf{2013}, \emph{7}, 7824--7832\relax
\mciteBstWouldAddEndPuncttrue
\mciteSetBstMidEndSepPunct{\mcitedefaultmidpunct}
{\mcitedefaultendpunct}{\mcitedefaultseppunct}\relax
\EndOfBibitem
\bibitem[Shibanuma \latin{et~al.}(2016)Shibanuma, Albella, and
  Maier]{shibanumaUnidirectionalLightScattering2016}
Shibanuma,~T.; Albella,~P.; Maier,~S.~A. Unidirectional Light Scattering with
  High Efficiency at Optical Frequencies Based on Low-Loss Dielectric
  Nanoantennas. \emph{Nanoscale} \textbf{2016}, \emph{8}, 14184--14192\relax
\mciteBstWouldAddEndPuncttrue
\mciteSetBstMidEndSepPunct{\mcitedefaultmidpunct}
{\mcitedefaultendpunct}{\mcitedefaultseppunct}\relax
\EndOfBibitem
\bibitem[van~de Groep \latin{et~al.}(2016)van~de Groep, Coenen, Mann, and
  Polman]{groep_direct_2016}
van~de Groep,~J.; Coenen,~T.; Mann,~S.~A.; Polman,~A. Direct Imaging of
  Hybridized Eigenmodes in Coupled Silicon Nanoparticles. \emph{Optica}
  \textbf{2016}, \emph{3}, 93--99\relax
\mciteBstWouldAddEndPuncttrue
\mciteSetBstMidEndSepPunct{\mcitedefaultmidpunct}
{\mcitedefaultendpunct}{\mcitedefaultseppunct}\relax
\EndOfBibitem
\bibitem[Tang and Cohen(2010)Tang, and Cohen]{tang_optical_2010}
Tang,~Y.; Cohen,~A.~E. Optical {{Chirality}} and {{Its Interaction}} with
  {{Matter}}. \emph{Physical Review Letters} \textbf{2010}, \emph{104},
  163901\relax
\mciteBstWouldAddEndPuncttrue
\mciteSetBstMidEndSepPunct{\mcitedefaultmidpunct}
{\mcitedefaultendpunct}{\mcitedefaultseppunct}\relax
\EndOfBibitem
\bibitem[Meinzer \latin{et~al.}(2013)Meinzer, Hendry, and
  Barnes]{meinzer_probing_2013}
Meinzer,~N.; Hendry,~E.; Barnes,~W.~L. Probing the Chiral Nature of
  Electromagnetic Fields Surrounding Plasmonic Nanostructures. \emph{Physical
  Review B} \textbf{2013}, \emph{88}, 041407\relax
\mciteBstWouldAddEndPuncttrue
\mciteSetBstMidEndSepPunct{\mcitedefaultmidpunct}
{\mcitedefaultendpunct}{\mcitedefaultseppunct}\relax
\EndOfBibitem
\bibitem[Zhao and Reinhard(2019)Zhao, and Reinhard]{zhao_switchable_2019}
Zhao,~X.; Reinhard,~B.~M. Switchable {{Chiroptical Hot}}-{{Spots}} in {{Silicon
  Nanodisk Dimers}}. \emph{ACS Photonics} \textbf{2019}, \emph{6},
  1981--1989\relax
\mciteBstWouldAddEndPuncttrue
\mciteSetBstMidEndSepPunct{\mcitedefaultmidpunct}
{\mcitedefaultendpunct}{\mcitedefaultseppunct}\relax
\EndOfBibitem
\bibitem[Wang \latin{et~al.}(2019)Wang, Fan, Luo, Cao, Wu, Zhang, Heller, and
  You]{wang_massive_2019}
Wang,~S.; Fan,~K.; Luo,~N.; Cao,~Y.; Wu,~F.; Zhang,~C.; Heller,~K.~A.; You,~L.
  Massive Computational Acceleration by Using Neural Networks to Emulate
  Mechanism-Based Biological Models. \emph{Nature Communications}
  \textbf{2019}, \emph{10}, 4354\relax
\mciteBstWouldAddEndPuncttrue
\mciteSetBstMidEndSepPunct{\mcitedefaultmidpunct}
{\mcitedefaultendpunct}{\mcitedefaultseppunct}\relax
\EndOfBibitem
\bibitem[Nadell \latin{et~al.}(2019)Nadell, Huang, Malof, and
  Padilla]{nadell_deep_2019}
Nadell,~C.~C.; Huang,~B.; Malof,~J.~M.; Padilla,~W.~J. Deep Learning for
  Accelerated All-Dielectric Metasurface Design. \emph{Optics Express}
  \textbf{2019}, \emph{27}, 27523--27535\relax
\mciteBstWouldAddEndPuncttrue
\mciteSetBstMidEndSepPunct{\mcitedefaultmidpunct}
{\mcitedefaultendpunct}{\mcitedefaultseppunct}\relax
\EndOfBibitem
\bibitem[Gallinet and Martin(2009)Gallinet, and
  Martin]{gallinetElectromagneticScatteringFinite2009}
Gallinet,~B.; Martin,~O. J.~F. Electromagnetic {{Scattering}} of {{Finite}} and
  {{Infinite 3D Lattices}} in {{Polarizable Backgrounds}}. \emph{Theoretical
  And Computational Nanophotonics (Tacona-Photonics 2009)} \textbf{2009},
  \emph{1176}, 63--65\relax
\mciteBstWouldAddEndPuncttrue
\mciteSetBstMidEndSepPunct{\mcitedefaultmidpunct}
{\mcitedefaultendpunct}{\mcitedefaultseppunct}\relax
\EndOfBibitem
\bibitem[Evlyukhin \latin{et~al.}(2010)Evlyukhin, Reinhardt, Seidel,
  Luk'yanchuk, and Chichkov]{evlyukhinOpticalResponseFeatures2010}
Evlyukhin,~A.~B.; Reinhardt,~C.; Seidel,~A.; Luk'yanchuk,~B.~S.;
  Chichkov,~B.~N. Optical Response Features of {{Si}}-Nanoparticle Arrays.
  \emph{Physical Review B} \textbf{2010}, \emph{82}, 045404\relax
\mciteBstWouldAddEndPuncttrue
\mciteSetBstMidEndSepPunct{\mcitedefaultmidpunct}
{\mcitedefaultendpunct}{\mcitedefaultseppunct}\relax
\EndOfBibitem
\bibitem[Hegde(2020)]{hegde_photonics_2020}
Hegde,~R.~S. Photonics {{Inverse Design}}: {{Pairing Deep Neural Networks With
  Evolutionary Algorithms}}. \emph{IEEE Journal of Selected Topics in Quantum
  Electronics} \textbf{2020}, \emph{26}, 1--8\relax
\mciteBstWouldAddEndPuncttrue
\mciteSetBstMidEndSepPunct{\mcitedefaultmidpunct}
{\mcitedefaultendpunct}{\mcitedefaultseppunct}\relax
\EndOfBibitem
\end{mcitethebibliography}

	\clearpage
	\section*{For Table of Contents Only}
	\includegraphics{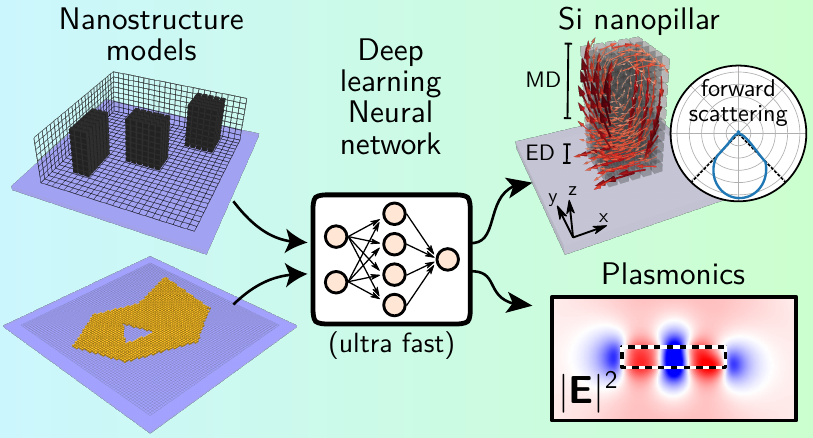}

\end{document}